**A Quick Study of Science Return from Direct Imaging Exoplanet Missions**
Detection and Characterization of Circumstellar Material with an AFTA or EXO-C/S CGI

Glenn Schneider (gschneider@as.arizona.edu)
Steward Observatory and the Department of Astronomy, The University of Arizona

ABSTRACT / EXECUTIVE SUMMARY

The capabilities of a high (~ $10^{-9}$ resel$^{-1}$) contrast, narrow-field, coronagraphic instrument (CGI) on a space-based AFTA-C or probe-class EXO-C/S mission, conceived to study the diversity of exoplanets now known to exist into stellar habitable zones, are particularly and importantly germane to symbiotic studies of the systems of circumstellar (CS) material from which planets have emerged and interact with throughout their lifetimes. The small particle populations in "disks" of co-orbiting materials can trace the presence of planets through dynamical interactions that perturb the spatial distribution of the light-scattering debris, detectable at optical wavelengths and resolvable with an AFTA-C or EXO-S/C CGI. Herein we: (1) present the science case to study the formation, evolution, architectures, diversity, and properties of the material in the planet-hosting regions of nearby stars, (2) discuss how a CGI under current conception can uniquely inform and contribute to those investigations, (3) consider the applicability of CGI anticipated performance for CS debris system (CDS) studies, (4) investigate, through AFTA CGI image simulations, the anticipated interpretive fidelity and metrical results from specific, representative, zodiacal debris disk observations, (5) comment on specific observational modes and methods germane to, and augmenting, CDS observations, (6) present, in detail, the case for augmenting the currently conceived CGI two-band Nyquist sampled (or better) imaging capability with a full linear-Stokes imaging polarimeter of great benefit in characterizing the material properties of CS dust (and exoplanet atmospheres, discussed in other studies).

CONTENTS

## 1. Introduction

Circumstellar (CS) material in "disks" orbiting stars are both the progenitors and the outcomes of planetary system formation processes. Huge strides have recently been made by NASA's *Kepler* mission, and ground-based radial velocity programs, toward a census of and statistical expectations for, exoplanets and exoplanetary systems. However, a deeper understanding of the complex nature of exoplanetary system formation, evolution, architectures, composition, and diversity, *inclusive of materials of all size scales co-orbiting with planets* (or escaping the systems) and *over a wide range of stellocentric distances*, must be informed by studying such material in the CS planet-hosting and interacting environments.

As known from *in situ* observations, our own solar system (SS) possesses a "two-component" CS debris system (CDS) with the outer, Edgeworth-Kuiper belt (EKB) region of cold dust co-located in the domain of the giant planets and beyond (to ~ 60 AU), and the inner zodiacal region of warm dust, in the zone of the terrestrial planets inclusive of the Earth, our Sun's habitable zone (HZ) and main asteroid belt. Whether such an architecture is typical in exo-planetary debris systems (i.e., what is "$\varepsilon_{dirt}$"?) is conjectural, as exo-zodiacal disks have yet to have been directly observed - though their existence is inferred from infrared to millimeter wavelength excess emission above the levels expected from their host stars ($L_{IR}/L_{star}$).

Our SS's CDS has an $L_{IR}/L_{star} \sim 10^{-7}$. If viewed from the "outside", at visible wavelengths, our SS's CDS would have a disk-integrated starlight-scattering fraction of $f_{disk}/f_{star} \sim 4\times10^{-8}$ (with spatially-resolved surface brightness (SB) then dependent on viewing geometry), and would be undetectable by about (or at least) an order of magnitude in spatially-resolved $f_{disk}/f_{star}$ contrast[1] with an AFTA coronagraphic instrument (CGI) under conception. However, all other factors being equal (though an unlikely scenario in detail if attempting to quantify a conceptual "$\varepsilon_{dirt}$"), as for $L_{IR}/L_{star}$, $f_{disk}/f_{star}$ is proportional to the total CDS dust mass. The spatial distribution of the CDS dust, however, will depend on the architecture of the planetary system (as well as stellar and environmental properties); so the conjecture of a SS-like CDS as archetypical is unlikely, in particular given the diversity in planetary system architectures as informed by *Kepler*. Such light-scattering CDSs may be probed into their zodiacal-belt regions with an AFTA or EXO-C/S CGI.

High-resolution optical-to-nIR imaging observations of (only) the outer, EKB-analog regions, of approximately two-dozen CDSs have been secured over the past two decades (mostly with the Hubble Space Telescope; *HST*). SB measures of the EKB components of these disks inform (through models constrained by thermal spectral energy distributions) of systemic dust masses many thousands of times that of our SS's EKB. These same observations also directly inform on a remarkable diversity in debris system morphologies and architectures in these outer, detectable, regions (e.g., see Fig. 2.1), though the inner, zodiacal-belt analog regions currently remain CS incognita due to instrumental sensitivity and inner working angle (IWA) limits (see Fig. 2.2).

## 2. Key Disk Science Questions Addressable with an AFTA or EXO-C/S CGI

The currently conceived AFTA CGI *raw* image contrast resel[-1] sensitivity ~ $10^{-8}$, with order of magnitude post-processing improvement at sub-arcsecond stellocentric IWAs (presumed throughout this study report), will, for the first time, uniquely enable spatially-resolved imaging into the zodiacal (and co-spatial HZ) warm dust regions of CDS's with postulated similarities, and diversities from our own SS's. This leads directly to the ability to explore some key

---

[1] In this study we define image contrast as the flux density contained within a resolution element at any stellocentric angular distance arising from the stellar point spread function (PSF) halo ratioed to the flux density within the central resolution element of the stellar PSF core unsuppressed by starlight suppression/coronagraphy.

questions in exoplanetary system science through the observations of CDS including:

- What are the levels of CS dust in, and exterior to, the habitable zones of exoplanetary systems that are posited to influence their evolution and character?
- Will dust in habitable zones interfere with planet-finding?
- How do dust sub-structures seen in CDSs trace the presence of (unseen) planets?
- What veneer is delivered to planetary atmospheres and surfaces by asteroids, comets, and other CS material?

And, extended to the inner regions also of transitional and protoplanetary disks:
- How do disks of protoplanetary materials evolve to make Solar System-like architectures?

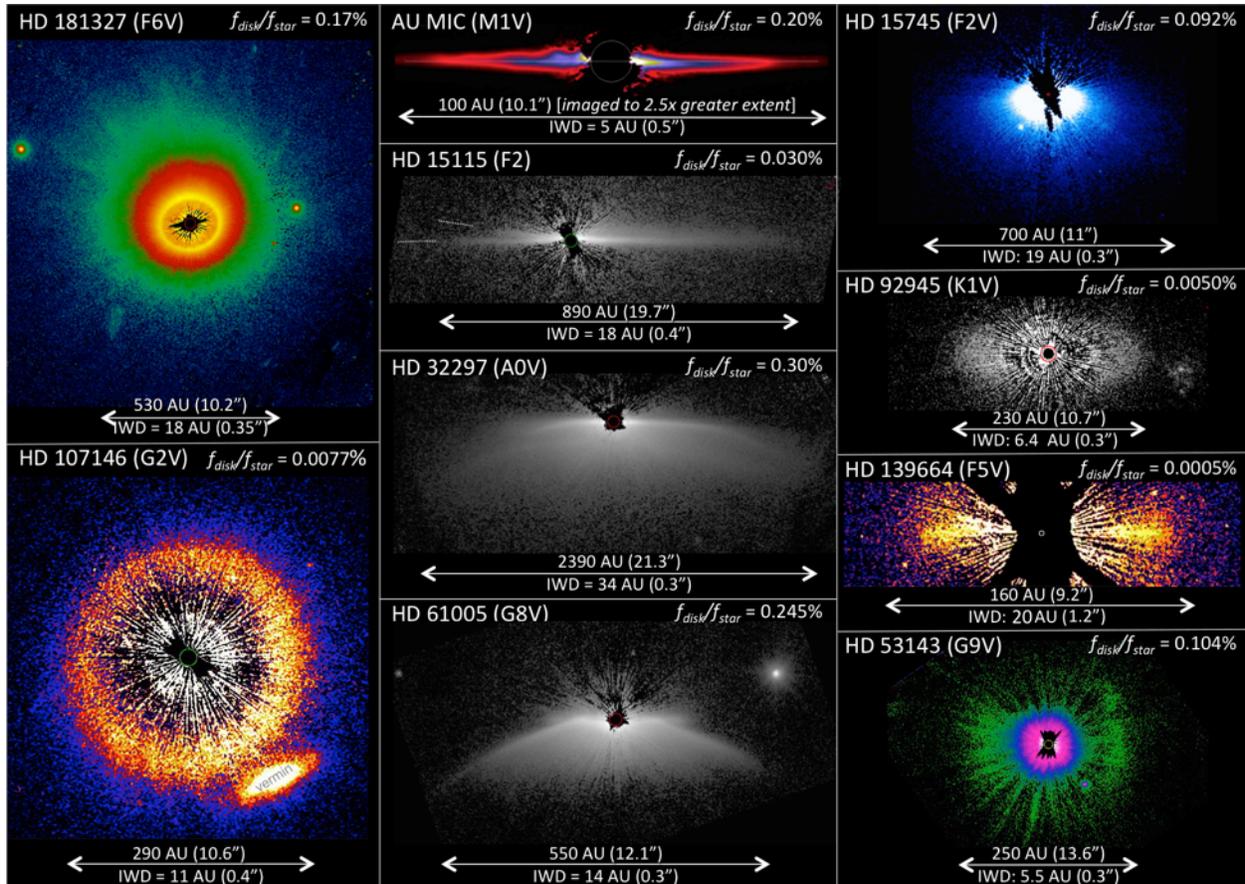

Fig 2.1. Ten (of 24 published to date) CDSs imaged in scattered light. This sample is homogeneously observed with *HST*/STIS PSF-template subtracted coronagraphy (reproduced from Schneider et al. 2014). These SB images underscore the diversity in outer disk morphologies and architectures. CDS optical scattering fractions (determined from these data) and stellocentric IWA achieved at *HST* contrast limits (see Fig 2.2) are confined to the EKB-analog regions imaged in these disks with corresponding physical distances (IWD: closest Inner Working Distance, not necessarily along the disk major axis) indicated. See Schneider et al. *ibid* for correlation of $f_{disk}/f_{star}$ with $L_{IR}/L_{star}$.

The answers to these and other related, highly compelling, and fundamental questions of astrophysical import may be forthcoming through direct imaging informing theoretical models imposing wavelength dependent constraints via filter photometry and/or IFU spectroscopy (both also benefitted with polarimetry; see § 8). The former are the baseline observational modes for a currently conceived AFTA CGI integrating a high-performance wavefront error control system and high-contrast coronagraph for starlight suppression. These proposed capabilities provide access to the inner regions of starlight-scattering CS disks in previously inaccessible observational domains of contrast and IWA required for studying the zodiacal-analog regions ripe for scientific exploitation.

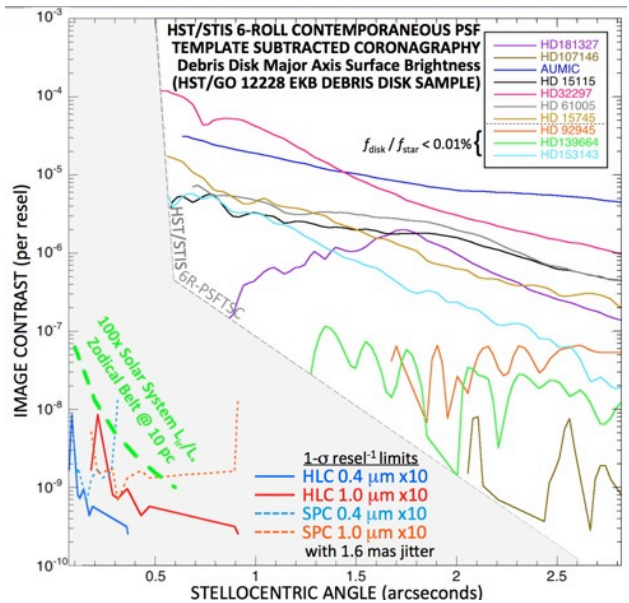

Fig 2.2. Top right (region above gray dashed line): Visible-light SB profiles of 10 (representative, but of highest image quality) of 24 CDSs imaged to date (see Fig 2.1 for corresponding images), measured along their disk major axes. The most favorable ($f_{disk}/f_{star}$ brightest) CDSs that are observable today are traced to stellocentric major axis IWAs ~ 0.5" - 0.7" (not smaller physical mask limits due to *HST* observational systematics) but only at contrast levels $\geq 5 \times 10^{-5}$ resel$^{-1}$ commensurate with massive EKB-analog disks in the stellocentric regions accessible. Fainter (and larger) EKB-analog only disks are observable at smaller contrast levels, but only at larger stellocentric distances (bottom right). Bottom left (below gray dashed line): Solid lines: CGI 3-σ resel$^{-1}$ detection limits for CS material in red and blue channels with anticipated x10 contrast augmentation via speckle calibration puts zodiacal disks (SS-analog with ~ 100x more dust in green) well within the grasp of either the AFTA Hybrid Lyot, or Shaped Pupil Coronagraphs (HLC, SPC).

## 3. Starlight-Scattering Circumstellar Dust in Exo-Planetary Systems.

How planetary systems, including our own, form and evolve is one of the most fundamental and compelling areas of inquiry in contemporary astrophysics. Over the last few decades, optical, IR and mm observations have shown that most stars are surrounded at birth by CS accretion disks of ISM-like primordial material. Separately, today, ~700 exoplanet candidates have been detected by RV techniques and > 5x as many by Kepler transit photometry. These planets are thought to originate in such disks, with Jovian-mass planets forming before CS accretion disks become gas-poor over time.

The evolutionary transformation of gas-rich disks to systems with small amounts of dust and little residual gas occurs over the first ~10 Myr of the star's lifetime, with warm dust rare after ~3 Myr. Much of the dust may then be tied up in the formation of smaller rocky or icy bodies that can interact with co-orbiting giant planets or smaller bodies, and collide, thus replenishing the CS environments (transformed into "debris disks") with second-generation material. Such collisional replenishment of dust-rich CS grains can continue throughout the system's lifetime. In our own solar system, even after 4.6 Gyr, this is evidenced by the presence of our zodiacal dust cloud, and asteroid and Kuiper belts. This dust-production paradigm was recently reinforced with *HST* imaging of the aftermath of the apparent collision of minor planet P/2010 A2 that released a wake of trailing debris, even as competing forces (e.g., Poynting- Robertson drag and radiation pressure "blow out") attempt to clear the circumsolar environment of such grains. The outcomes of such interactions will analogously affect the composition and spatial distribution of CS debris on spatial scales large compared to planets.

Interplanetary dust in CDSs not only absorbs stellar radiation that is then re-radiated at thermal IR to mm wavelengths, but also scatters it at visible wavelengths. Even cold grains far stars are illuminated by starlight. Scattered-light images in the "cold dust" regions, such as those imaged with *HST* (e.g., Fig 2.1), have thus far provided the greatest insights because they have traced dust at a wide range of stellocentric distances at high spatial resolution. For example, images enabled with optical-to-nIR coronagraphy at wavelengths from 0.5 – 2 μm of HR 4796A show a narrow EKB-analog region ring that, by analogy with the (larger) Fomalhaut debris ring, may be shepherded (and dynamically sculpted) by unseen planets (Schneider et al. 1999, Debes et al. 2008, Schneider et al. 2009); sub-mm imaging with ALMA symbiotically traces the large grain and planetesimal populations (Boley et al. 2014; ESO14), while a CGI reveals the small grains. The CS disk associated with the Herbig AeBe star HD 141596 reveals spiral and/or arc-like sub-structures in its observable EKB-analog region with an inner gap indicative of the dynamical effects of an embedded giant planet (e.g., Weinberger et al. 1999, Clampin et al. 2003, Ardila et al. 2005; Wyatt 2005). Such features are proving not uncommon in spatially-resolved scattered-light images of CDS and also in outer regions of those transitional disks that are now within the grasp of some ground-based AO augmented instruments. *No instrument yet built, however, has had the inner working angle and imaging contrast to detect and spatially resolve zodiacal disks containing CS habitable zones* where temperatures allow liquid water to remain on a planet's surface (e.g., Kasting et al. 1993); that area in which the Earth resides in our own solar system. Diffuse dust in this inner region of our solar system produces the zodiacal light, which we see extending in the plane of our solar system, i.e., our own circumsolar disk of tenuous dusty debris. More massive examples of extra-solar zodiacal disk analogs would be observable, and resolvable, with an AFTA or EXO-C/S CGI.

Planets set the locations where planetesimal belts and their debris can stably persist. The maintenance of the inner edge of our Kuiper Belt by Neptune (Liou & Zook 1999), the structure of Fomalhaut's CS debris ring and its co-orbiting planet (Kalas et al. 2005, 2008), and the inclination of β Pic b to its "warped" (or two component) debris disk are among the very few directly observable examples of dynamical links between planets and CDSs. Spatially resolved imaging of the CS debris in the warm inner zodiacal dust regions of nearby stars will reveal the now-elusive (SS analog) terrestrial planet and habitable zones of dusty exoplanetary systems and provide direct evidence for asteroid belts, comets and the unseen parent bodies responsible for the light-scattering CS debris. An AFTA or EXO-C/S CGI will uniquely provide detailed information about these CDSs and their architectures even if individual planets remain elusive.

## 4. Imaging CS Material – $L_{IR}/L_{star}$ "Expectations"

The inexorable link between CS dust, (exo)-planets, systemic evolution, and their morphologies and architectures was first suggested as a general proposition by Kant (1755) and Laplace (1796). Two centuries would pass, however, before technologies matured enabling the first (indirect) evidence of such CS material, secured photometrically, as thermal infrared emission above stellar photospheric brightness levels. These initial detections (beginning with Vega), ascertained with NASA's *IRAS* satellite (Aumann, 1984), were interpreted as arising from re-radiation of absorbed stellar flux by debris particles (possibly) in Keplerian orbits. This conjecture was viscerally demonstrated as correct with the first spatially-resolved image of an *IRAS*-detected CDS - the prototypical light-scattering CS debris disk of β Pictoris (Smith and Terrile 1984) with a large fractional IR luminosity of the dust in the system compared to the star: $L_{IR}/L_{star} \approx 2 \times 10^{-3}$.

Fractional IR luminosity is not directly a measure of the dust mass in a CDS as $L_{IR}/L_{star}$ is dependent upon the particle properties (e.g., size, composition, structure), as well as stellocentric distance. $L_{IR}/L_{star}$, however, is an easily measurable quantity and is often used as a first-order proxy for the amount of dust in the system (Decin et al. 2003). In an optically thin medium, as may be expected for dust in a CDS, $L_{IR}/L_{star}$ is proportional to the systemic dust mass, and so is a commonly used metric for inference of material visibility. $L_{IR}/L_{star}$, as such, has frequently used as a primary criterion for target selections in disk imaging surveys (with a simplifying, but often unrealistic, presumption with all other things being equal that more dust presenting a higher surface density for light-scattering results in concomitantly higher detectability). β Pictoris is among the $L_{IR}/L_{star}$ brightest CDSs known. The seminal ground-based scattered-light imaging of the β Pictoris disk was possible due not only to its very large $L_{IR}/L_{star}$, but, in combination its fortuitous (edge-on) viewing geometry (enhancing its visibility), and very large angular extent due to its serendipitous proximity to the Earth making it much less of a "contrast challenge" than most other CDSs.

Over time, hundreds of such posited dust-bearing systems were identified through their thermal excesses as measured with *IRAS*. Direct imaging examples of even the $L_{IR}/L_{star}$ brightest CDSs (other than β Pic), however, remained elusive until the era of space-based coronagraphy with *HST* (e.g., Schneider et al. 1999, Weinberger et al. 1999, Schneider. et al 2001). Even then, *HST* surveys for light-scattering CDSs with near-IR and optical coronagraphy of systems with "conservative"[2] $L_{IR}/L_{star} \geq \sim 10^{-4}$ selections were returning yields (imaging detections) of only ~15%. This is indicative of the fact that $L_{IR}/L_{star}$ alone is not a strong predictor of optical/nIR scattered-light SB in the absence of constraints otherwise informed by systemic geometries and particle properties. *HST* coronagraphy has provided access to the outer "cold" dust (Kuiper-belt analog) regions of CDSs, mapping complex and spatially extended starlight-scattering debris structures (for examples, see Fig 2.1). The inner "warm" (zodiacal-belt analog) dust regions betrayed by their stellar thermal IR excesses (into habitable zones), however, remain today undetectable in scattered light, but will be within the grasp of an AFTA or EXO-C/S CGI.

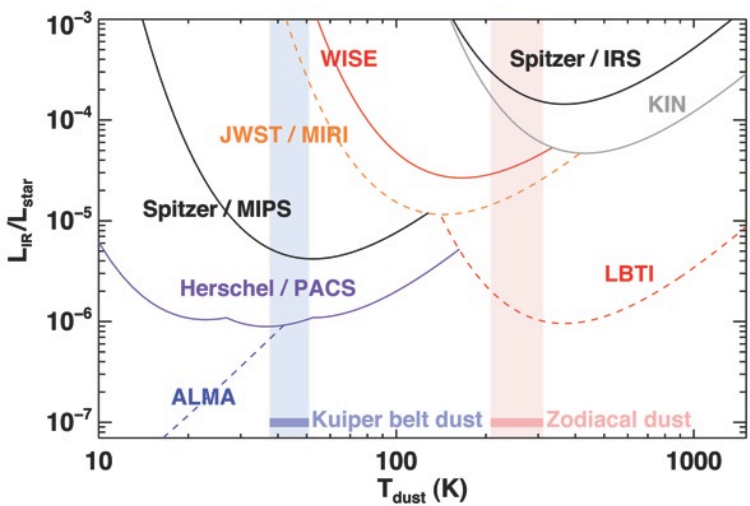

Fig 4.1. Since *IRAS*, follow-on space-IR facilities (solid lines, plus the Keck Interferometer; KIN) have provided additional candidates for scattered-light imaging of CDSs to their $L_{IR}/L_{star}$ with peak wavelength-dependent sensitivities over a wide range of dust temperatures (and thus stellocentric distances for homogeneous blackbody grains). For reference, our solar system's Kuiper and zodiacal belts are shown and are an order of magnitude below the sensitivity limits of Herschel/PACS and LBTI (anticipated), respectively. From Roberge et al. 2012.

IR facilities subsequent to *IRAS*, with improved sensitivities and capabilities providing additional and smaller $L_{IR}/L_{star}$ detections (see Fig 4.1), have informed new and conceived follow-on scattered-light imaging surveys. However, the current focus primarily remains on the higher probability imaging detections with $L_{IR}/L_{star}$ bright CDSs for which "cold" dust temperatures suggest the possible presence of light-scattering debris at stellocentric distances within the working angle reach of current high-contrast imaging systems.

Recent visible-light observations of a sample of the outer regions of ten CDSs observed with *HST* coronagraphic sensitivities show a remarkable diversity in disk morphologies and sub-structures from 0.3" (smallest *HST* coronagraphic IWA) outward, beyond those anticipated from simple geometric projection and $r^{-2}$ attenuation of illuminating starlight. Such spatially resolved scattered-light images of CS debris in exoplanetary systems constrain the physical properties and orbits of the dust particles in these systems in their EKB-analog regions thus-far explored. They also inform on co-orbiting (but unseen) planets, the systemic architectures, and forces perturbing the starlight-scattering circumstellar material.

*HST*, and recent advances in ground-based EXAO high-contrast imaging systems, have greatly informed on the possibilities for the small number of systems thus far imaged in their outer, Kupier-belt analog, regions explorable to date. The inner regions of these systems, analogous to our SS's zodiacal belt and HZ are beyond those capabilities. The capabilities afforded by a proposed AFTA CGI with $10^{-9}$ augmented resel$^{-1}$ contrast at sub-arcsecond IWAs can break this paradigm of exclusivity in stellocentric angular extent.

---

[2] A high-fidelity analog to our own two-component (Kuiper + zodiacal belt region) circumsolar debris system has an $L_{IR}/L_{star} \approx 10^{-7}$, so as a proxy for dust mass, $L_{IR}/L_{star} = 10^{-4}$ is ≈ 1000x our solar system (SS). Our SS viewed face-on from 10 pc would have an optical light scattering fraction, $f_{disk}/f_{star} \approx 4 \times 10^{-8}$. Correlation with scattered light surface brightness depends on dust properties, spatial distribution, and line-of-sight inclination, but for a true SS analog CDS with 1000x dust mass one would expect $f_{disk}/f_{star} \approx 4 \times 10^{-5}$ which is approximately at the limit approaching *HST* sub-arcsecond IWA capabilities at visible wavelengths (with STIS); e.g., see Fig 5.2 lower right panel.

## 5. CGI Zodiacal-Analog Disk (ZAD) Targets vis-à-vis Science Goals

A $10^{-9}$ resel$^{-1}$ contrast CGI will enable spatially-resolved imaging of ZADs into stellar HZs, where liquid water, considered a vital ingredient for life, can exist on planetary surfaces. With CGI augmented image contrasts ~ $10^{-9}$ resel$^{-1}$ in HZs, the selection of targets for CDS imaging in ZADs can be unbiased by stellar ages for nearby systems with sufficiently large amounts of warm dust ($L_{IR}/L_{STAR} \geq$ few × 10 of our SS, with $T_{dust}$ a few hundred Kelvins).

*What are the levels, and nature, of the dust in the HZs of exoplanetary debris systems?* The CGI spatial resolution enables mapping of the two-dimensional spatial distribution of ZAD dust of sufficient scattered-light SB (see § 7 for AFTA CGI examples). The amount of scattered light seen is a proxy for the richness of the planetesimal belts and their degree of gravitational stirring; it is an indirect indication of the level of small particle bombardment that might be experienced by terrestrial planets in these systems. Radially differentiated image structure (SB, color, and polarization with a measurement-capable CGI; see § 8) informs on stellocentric segregation of particles with temperature dependent properties. Azimuthal asymmetries in dust SB distributions beyond those attributable to scattering phase functions inform on possible perturbations (resonances, sculpting, pericentric offsets of dust features, etc.) due to co-orbiting planets that may or may not be detectable (e.g., see § 7.4), but can be modeled to constrain planetary orbits and masses.

*Will dust in HZs interfere with planet‑finding?* The amount of dust in the HZ is critical in determining the best strategies to later image Earth-like planets around other stars since light scattered by CS dust is the main source of astrophysical noise in detecting such faint, embedded, point sources. The typical masses measured for protoplanetary disks appear lower than that needed to form the SS planets (Andrews & Williams 2005). The amount of debris disk emission is proportional to, and decays from, the mass in the corresponding protoplanetary disks (Wyatt et al. 2007b). Putting these results together suggests that the absence of debris dust in an evolved system may indicate that it had a low-mass protoplanetary disk to begin with. It is therefore at least conceivable that by (preferentially) targeting stars without debris dust, future exoplanet imaging missions may be selecting targets unlikely to have had sufficient initial mass for rich planetary systems. That is, optimal strategies for imaging Earth-like planets in HZs may have to balance competing considerations about CS debris. Doing so requires a deep understanding of the prevalence and evolution of debris dust, an understanding that simply cannot be achieved with the limited number of systems currently imaged, and imaged only in their EKB-analog regions. AFTA CGI enabled ZAD images, over a wide range of stellar types and ages, will form the foundation of studies of disk behavior as functions of stellar and environmental properties.

Primary samples for consideration in building an AFTA CGI disk-science DRM include: (1) the nearby stars: (a) distance limited, and (b) additionally with radial velocity (RV) detected exoplanets, (2) CDSs with previously imaged EKB-analog region disks, (3) light-scattering CDSs candidates suggested from newly-identified long wavelength excesses and imaging, (4) young stars hosting known protoplanetary disks (e.g., ESO14). We comment further on (1) - (3).

### 5.1. Nearby FGK Stars

In Fig 5.1 we show the stellocentric reach of the AFTA CGI's proposed *B* and *Z* filter bands

(see Table 1) w.r.t. ZADs extending to stellar-luminosity scalable outer zodiacal disk radii of 6 AU, and their corresponding HZs[3] for F, G, and K stars in a distance-limited sample with d < 25 pc. The target stars plotted in Fig 5.1, with zodiacal-analog belt regions accessible to an AFTA CGI (here within the HLC stellocentric working angle zones), do not represent an $L_{IR}/L_{STAR}$ selected (biased) sample – but are designed to probe systems with putative dust amounts below those currently detectable from warm dust surveys that, for some targets, LBTI may soon inform. A distance-limited systematic survey, extending in many cases to the stellar HZs, will reveal the levels of zodiacal light present in CDSs imaged (to CGI contrast-limited detection sensitivities).

Stars in this sample hosting RV detected extrasolar planets will be of particular interest as "high priority" targets to better understand the connections between *a prior* known exoplanets and the CS material with which they are posited to interact in CDSs.

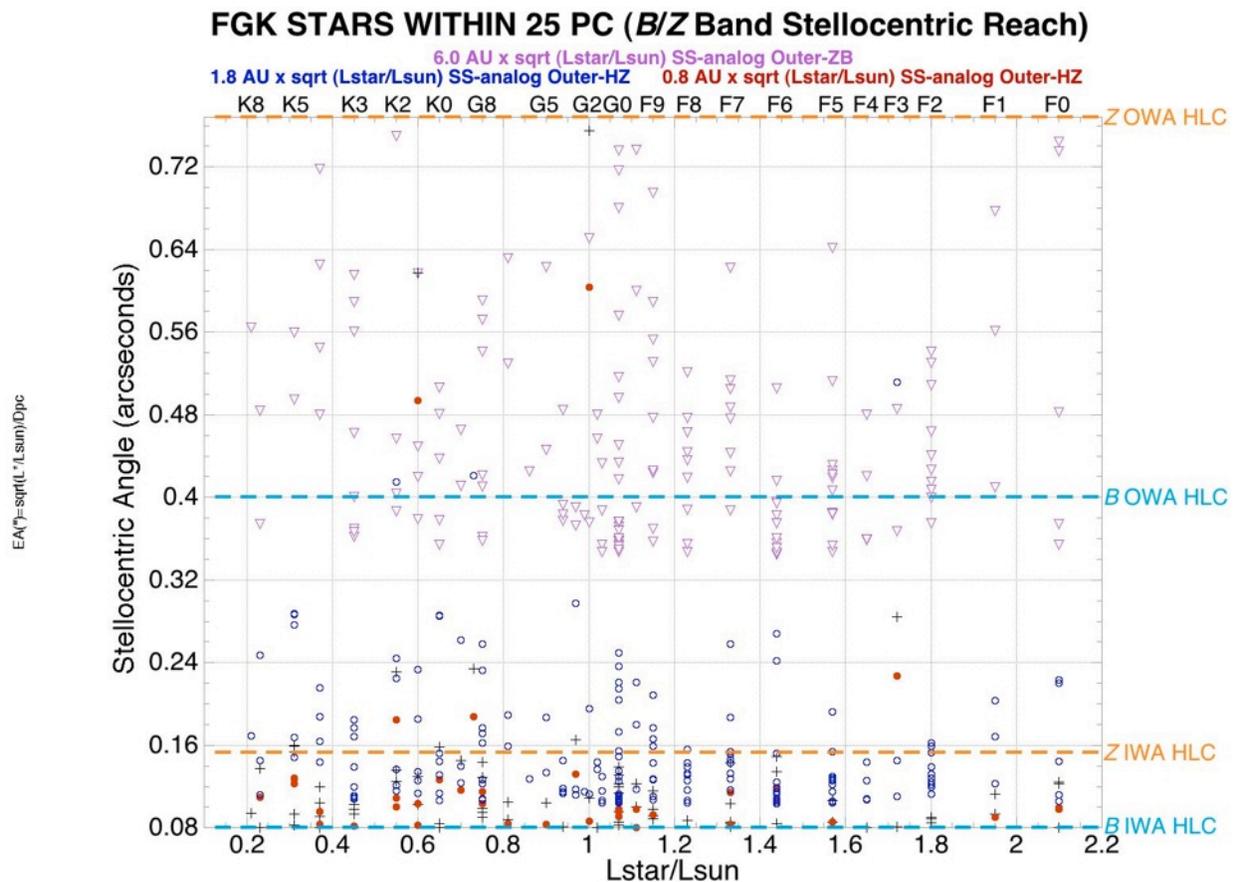

Fig 5.1. Nearby stars with zodiacal-belt analog regions accessible to AFTA CGI (HLC).

**5.2. CDSs with Previously Imaged EKB-Analog Region Disks**

The EKB-regions of approximately two-dozen CDSs have been imaged to date (mostly with *HST* coronagraphy), all with starlight-scattering fractions suggesting massive amounts of dust in the outer parts of those systems compared to our SS. We illustrate in Fig 5.2 with one such "ring-like" CDS about the close-solar analog star HD 107146 (G2V; 27 pc; with an estimated age ~ 80

---

[3] Herein we adopt HZ limits as 0.8 to 1.8 AU scaled by sqrt($L_{star}/L_{sun}$) and, for consistency, scale the outer distance of the zodiacal dust commensurately based loosely on our SS - though the actual dust distribution will depend upon the individual exoplanetary system architectures, hence our consideration of stars here out to 25 pc.

– 200 My). Its massive, well-revealed, EKB-analog region debris ring is 2000 times brighter in $f_{disk}/f_{star}$ than our SS's seen with a comparable viewing geometry. But detection of dust at even at twice the outer distance of its ZAD region (beyond the edge of the *HST* coronagraphic mask at 0.44") is contrast-limited at $10^{-3}$ resel$^{-1}$.

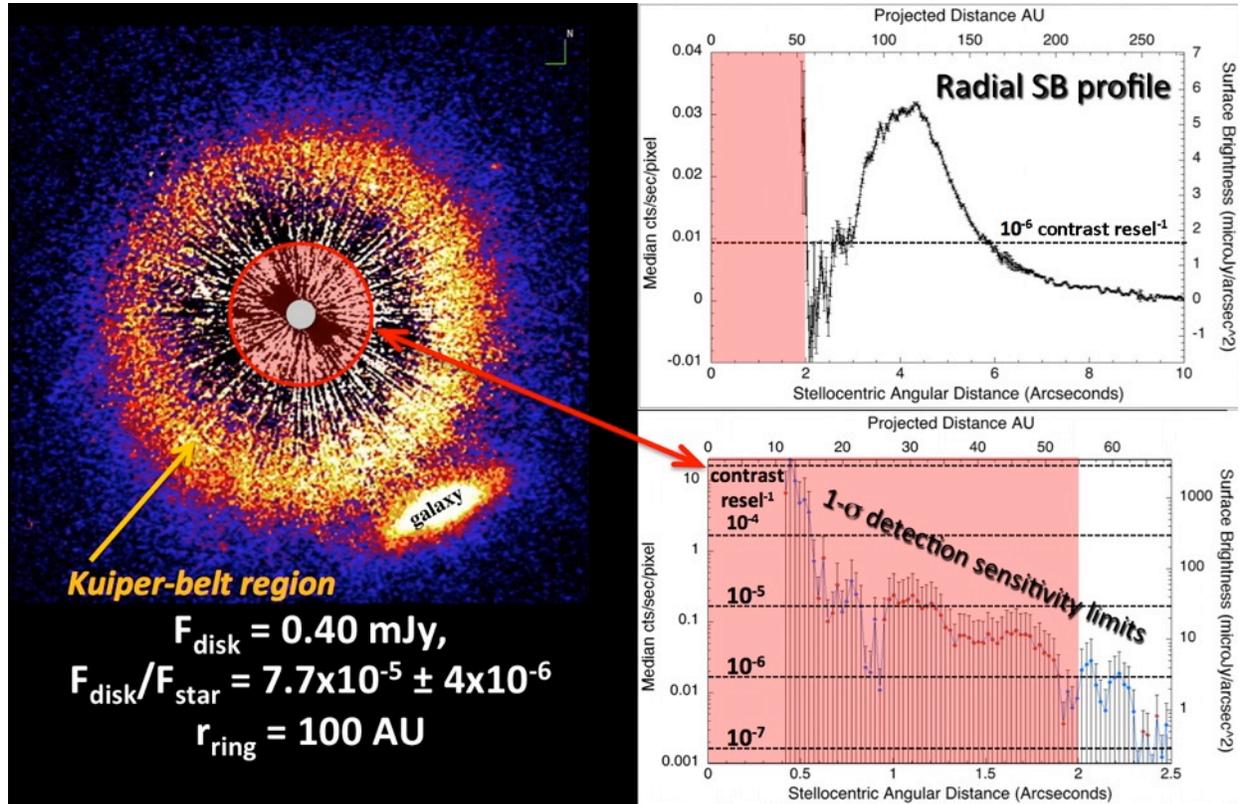

Fig 5.2. *HST* optical scattered-light image of the HD 107146 EKB-analog region debris ring and instrumental limits precluding dust detection in the interior part of the CDS (adapted from Schneider et al. 2014).

Two of the currently known light-scattering CDSs in this "legacy" sample, β Pictoris (A5V) and Fomalhaut (A3V), have co-orbiting planets that have been revealed by imaging (in part due to their fortuitous proximity to Earth); in the (exceptional) case of β Pic b, orbiting with $a \sim 8$ AU within its stellar-luminosity scaled outer zodiacal-belt analog radius. Its massive debris system, however, bears little resemblance to our own SS (e.g., for holistic views of the β Pic system see Apai et al. 2014) and some of its details are self-obscured due to its edge-on viewing geometry. The architectures of the zodiacal-belt analog regions of this otherwise well-studied legacy sample of CDSs with well-resolved EKB-analog regions remains unknown and ripe for exploration with AFTA or EXO-C/S CGI imaging. This is particularly the case for those known with intermediate to face-on viewing geometries amenable to self-unobscured observations of their innermost regions.

### 5.3. Light-Scattering CDS (Candidates) Detected at Longer Wavelengths

Most light-scattering CDSs imaged to date were (originally) suggested from *IRAS* measured $L_{IR}/L_{STAR}$ excesses. Subsequent samples with improved sensitivities and to dust of different temperatures (see Fig 4.1,) have (and continue) to suggest new CDS targets in both their outer (cold dust) EKB-analog, and inner (warm dust) ZAD regions. Some of these cold, EKB-region, dust structures (i.e., detected with Spitzer/MIPS ~ $5 \times 10^{-6}$, and Herschel/PACS ~ $1 \times 10^{-6}$, $L_{IR}/L_{STAR}$ sensitivity limits) have been spatially resolved at thermal IR wavelengths (e.g., Booth et al. 2013). Warm ZAD dust, detectable with Spitzer/IRS and KIN (both with ~ $10^{-4}$ $L_{IR}/L_{STAR}$ sensitivity limits, and soon anticipated two order of magnitude improved sensitivity with LBTI), is not yet detectable in scattered-light, but would be with an AFTA CGI. For high-yield (and high SNR) targeted ZAD imaging (as opposed to an unbiased surveys for more tenuous dust) such debris dust survey results would importantly inform "high priority" CGI target selections.

Until recently, the inner ~ 20 AU of many IR identified CDSs were thought to be relatively devoid of dust. However, it is now known from near-IR interferometric (e.g., di Folco et al. 2007; Stock et al. 2010) and mid-IR photometric detections (e.g., Stapelfeldt et al. 2004; Wyatt et al. 2005b) that at least some CDSs possess dust in close proximity to their central stars (e.g., Beichman et al. 2005; Song et al. 2005; Hines et al. 2006; Weinberger et al. 2011). Indeed, there is growing evidence from IR photometry for *multiple belts of interplanetary dust inside the planetary zones* (Backman et al. 2009; *c.f.* their Fig. 9). The origin of this terrestrial temperature dust is uncertain, though the rapid dust depletion timescales at these orbital distances imply that such grains must be transient in nature (Wyatt et al. 2007a). It is also not yet known whether this dust originates in planetesimals that have been dynamically scattered through some kind of systemic (possibly extrinsic) instability, or from recent collisions between massive protoplanets.

## 6. The Contrast-Limited Detectability of Spatially Resolved Scattered-Light Disks

The detectability of exoplanets against the glare of their host stars, as spatially unresolved point sources, is contrast-limited by the ability of any instrumental starlight suppression system to stabilize, reduce, and in post-processing characterize and calibrate, the intensity of instrumentally scattered and diffracted starlight in the stellocentric regions of interest. The instrumental inner-to-outer working angle range and corresponding (azimuthally averaged) stellocentric planet-to-starlight contrast ratio goals and requirements, enabled by a $\leq 10^{-9}$ contrast sub-arcsecond CGI, have been driven with priority for exoplanet direct detection and characterization into the HZs of nearby planet-hosting systems; see parallel "Quick Study" reports by Burrows, Marley and Renye. A "true" SS-analog zodiacal debris belt, with material orbiting within a few AU around a solar "twin" host-star, at a distance of $\approx$ 10 pc from the Earth, would appear within a stellocentric annulus of outer radius $\approx$ 0.6". This is within the CGI high-contrast stellocentric field (see Table 1). The requisite enabling instrumental attributes in these domains of observability for exoplanets (i.e., augmented image contrast with post-processing ~ $10^{-9}$ pixel$^{-1}$ at stellocentric angles ~ 0.1" – 1.0" at visible wavelengths with Nyquist or better pixel sampling) are also immediately applicable and are particularly germane for uniquely informing on the dust content in CS planet hosting environments, i.e., in CS protoplanetary, transition, and debris systems.

Detection sensitivity metrics for exoplanets, in the contrast-limed domain, are appropriately expressed for spatially unresolved point sources where most of the planet-light is contained within a single pixel or diffraction-limited resolution element (resel). Discrimination between residual CGI-induced point-like optical artifacts (before or after PSF subtraction/calibration) and true planets by various methods may disambiguate these signals that present themselves on similar, if not identical, spatial scales. Unlike planets, CDSs are spatially resolvable at optical wavelengths in an AFTA (or EXO probe) telescope. In many cases CDSs are anticipated to completely fill the high-contrast, WFE controlled, field-of-view (and extend well beyond its extrema) with potentially observable structures and sub-structures that inform on the distribution and properties of light-scattering material in their inner, zodiacal-belt analog regions. For CDSs, the total (area integrated) flux density of the starlight scattered by the disk particles into the observer's line-of-sight, $f_{disk}$, is (in principal) a measurable quantity analogous (in the exoplanet case) to the total planet light. This differs in CDSs, however, in being spatially distributed over a large area often with morphological complexity attributable to both the underlying physical processes responsible for sculpting the CS dust, and the happenstance of viewing geometry. Never-the-less, the global ratio of the total disk light to that of its host star, $f_{disk}/f_{star}$ (i.e., the total dust "scattering fraction"), is a useful order-of-magnitude metric to gauge the overall relative brightness (and potential detectability) of CDSs modulo the spatial distribution of the dust and its sky-plane projection.

The contrast-limited detectability of azimuthally and radially localized debris structures and substructures on various (resel and larger) correlated spatial scales, however, depends upon the SB distribution of the light-scattering dust. For the approximately two-dozen CDSs thus far observed, this can vary by orders of magnitude spatially within any CDS, or from CDS to CDS, even if compensated for the $r^{-2}$ falloff of the stellar radiation field in the plane of the disk. Viewing geometry alone (e.g., face-on vs. edge-on) can result in order of magnitude differences in both $f_{disk}/f_{star}$ and localized SB in structured disks, but details of (*a priori* unknown) dust density distributions can be responsible for larger dispersions in localized SBs and image contrast resel$^{-1}$.

The detectability (visibility) of disk structure and sub-structures is conflated by residual artifacts (correlated image structures) in the incompletely suppressed stellar PSF halo before or after PSF calibration/subtraction. The spatial scale(s) of the correlated residuals can create artifacts that, without identification or mitigation, can (in some cases) be indistinguishable from true disk features.

### 6.1 AFTA CGI Stellocentric High Contrast Field

Herein we assume an AFTA CGI high-contrast stellocentric field with *raw* image contrasts ~ $10^{-8}$ at a stellocentric angular distance of ~ 4 $\lambda$/D (D = 2.4 m) and $\geq$ x10 contrast improvement through PSF post-processing by speckle calibration or other methods. Similar (or better) contrast performance is expected for EXO-C/S. The AFTA HLC and SPC designs have similar wavelength-dependent inner-to-outer working angle ranges (see Table 1, with a recently conceived larger "disk mask" under consideration) and resel$^{-1}$ image contrasts as shown in Fig. 2.2 (scaled to the extrema of the wavelength range for filter band imaging or exoplanet IFU spectroscopy).

### 6.2. CGI Spectral Filters for Disk Imaging

For disk imaging, we consider (a minimum of) two filter bands near the blue and red extremes of the CGI spectral range to provide a factor of ~ x2 in wavelength for maximum leverage to ascertain an optical color index for constraining particle properties in stellocentric zones commonly sampled in both bands.

**Optical Color and Grain Types**. Over the CGI wavelength range:
(1) large silicate grains are neutral reflectors of stellar light;
(2) organic-rich grains, such as those apparently observed in the HR 4796A disk (Debes et al. 2008), are red and would be ~ 50% brighter at 400 – 500 nm than at 800 – 900 nm;
(3) only very small grains, such as those recently produced in high velocity collisions, produce a blue, Rayleigh scattered signature.

For representative dependencies of disk absorption and scattering efficiencies, and directionally preferential scattering phase function asymmetry factors (as well as degree of polarization) over the CGI ~ 2x wavelength range, see Fig 8.2.

We canonically consider a "*B*" filter with $\lambda_{central}$ = 465 nm, and a "*Z*" filter with $\lambda_{central}$ = 890 nm, both with spectral bandwidths R = FWHM(nm) / $\lambda_{central}$ of ~ 15 – 20%. The central wavelengths are not critical, as they are not designed to match up with diagnostic spectral features (e.g., absorption bands that are largely absent at optical wavelengths). Strong intra-band optical color gradients with R $\leq$ 20% are not expected, so narrower bandwidth filters for spectral color diagnosis are unnecessary and broadband R = 15 – 20% provides greater sensitivity to dust-scattered continuum starlight. These particular bands satisfy the ~ x2 wavelength range criterion and are commensurate with the "F465W" and "F890W" filters proposed for exoplanet imaging with similar bandwidths (the latter used in conjunction with the IFU for exoplanet studies).

The *B* filter band provides the highest spatial resolution (e.g., 1 resel = 0.49 AU at 10 pc) critical to resolving and mapping disk sub-structures and asymmetries indicative of dynamical interactions with (possibly unseen) exoplanets. The *B* filter band also provides the smallest inner working distances (e.g., ~ 0.8 AU at 10 pc with the HLC) key to reaching into the inner parts of SS-analog zodiacal disks and stellar HZs. The *Z* filter provides the largest outer working angle (OWA) reaching in many cases beyond the outer regions of SS-analog zodiacal disks (e.g., 7.6 AU at 10 pc with AFTA HLC, or ~ 18 AU at 10 pc for a conceptualized SPC "disk mask").

The inner-to-outer working angle range of the CGI will scale with wavelength. Having an intermediate region of stellocentric spatial overlap in the *B* and *Z* bands is important to look for telltale signs by radial color changes (and also polarization, if instrumentally capable; see § 8) of temperature segregated grain populations and/or ice-lines. With these filters, the stellocentric angular ranges of overlap in *B* and *Z* band resels is from the *Z*(IWA) to the *B*(OWA) to search for such color differentiated particle sizes and properties. For a face-on disk (worse case for this purpose) this probes a small, but useful, stellocentric annular zone - better (larger) with HLC than SPC. The physical range in stellocentric distance (AU) will increase for (most) disks seen in stochastic sky-plane projection with systemic inclinations - most usefully for disks seen in (most likely) intermediate inclinations between face and edge on viewing orientations. Importantly, the SPC is (additionally) evaluating a larger IWA-to-OWA "disk mask" which could be extremely valuable to push further out in stellocentric angle into also the inner regions of EKB-analog disks CDS closer to Earth, and make accessible a larger range of stellocentric distances.

Table 1 – Currently Conceived AFTA CGI Masks and Spectral Filters for Disk Imaging

|  | $\lambda/d$ |  | *B* (465 nm) |  | *Z* (890 nm) |  | Overlap B / Z |
|---|---|---|---|---|---|---|---|
|  | IWA | OWA | IWA | **OWA** | **IWA** | OWA | resels |
| HLC | 2.0 | 10.0 | 0.080" | **0.400"** | **0.153"** | 0.765" | 5.1 / 2.6 |
| SPC | 2.7 | 9.1 | 0.108" | **0.364"** | **0.207"** | 0.696" | 3.3 / 1.7 |
| *SPC/Disk Mask* | *6.5* | *24* | *0.250"* | ***0.959"*** | ***0.497"*** | *1.836"* | *9.5 / 5.0* |

## 7. Representative Disk Image Simulations with an AFTA CGI

In this section we present results from debris disk image simulations for an AFTA CGI using a shaped pupil coronagraph with an image plane mask[4] providing an IWA of 4.0 $\lambda/D$ and OWA of 12.4 $\lambda/D$ (corresponding to 187 to 583 mas at $\lambda$ = 550 nm). These scattered light image simulations are at a central wavelength of 550 nm with a spectral bandwidth R = 10%; a broader filter (suggested up to R ~ 20%) would image disks more efficiently. The simulated images over-sample the 550 nm PSF with 14.4 mas square pixels (i.e., 3.3 pixels per $\lambda/D$ at 550 nm). Noiseless disk models are generated using the ZODIPIC[5] IDL code (M. Kuchner and C. Stark, priv. comm.) assuming $T_{dust} \sim r^{-\Delta} \times L_{star}^{-\Delta/2}$ and $\Delta = 0.467$ (from the DIRBE model; Kelsall et al. 1998). The noiseless disk models[6] are convolved with the SPC PSF for the same filter band (see § 7.5). The *raw* stellocentric contrast field ("speckle" file provided by J. Krist) is attenuated by x10 as a presumed minimum planned/achievable with speckle calibration (subtraction) in post-processing. Photon noise from both the disk and the *raw* (unattenuated) speckles are added with integration time of 10 hours. An end-to-end system throughput efficiency of 35% is assumed.

---

[4] This "wedge" mask differs in working angle extrema from the current Table 1 baseline, and provides only an azimuthally restricted, ≈ 56°, "bow-tie" dark zone; it was adopted for simulations in this study for the timely availability of performance and related reference/calibration data with then a 360° field synthesized by replication.
[5] http://asd.gsfc.nasa.gov/Marc.Kuchner/zodipic.2.1.tar
[6] Including a scattered-light optical phase function from Hong et al. 1985 in § 7.4, also see footnote 9.

### 7.1. A Dynamically "Featureless" Disk – a 10 h visit on an close-solar analog

The first of these simulated images, shown in Fig. 7.1, is for a hypothetical SS-like, optically thin, dust disk circumscribing the planet-hosting star 47 Uma (G1V; 14.2 pc; V = 5.04; age ~ 6 Gy, Saffe et al. 2005), but with no model planets implanted. The model disk has 30 "zodis" of SS-like dust[7] with a power law particle density distribution $\propto (r/r_0)^{-\alpha}$ with $\alpha = 1.34$ as suggested by the DIRBE model of our SS (Kelsall et al. 1998). The disk is "featureless" in that its exhibits no central clearing (e.g., by planets, or dust sublimation exterior to the IWA limits of the coronagraphic mask at 2.7 AU), with no deviations from a continuous radial power-law profile nor azimuthal asymmetries that might arise from co-orbiting planets. The disk is viewed with a sky-plane projected inclination of 60° from face-on resulting in an $r^{-2.27}$ power law index SB distribution along the disk major axis. The major axis of the disk is oriented 45° w.r.t. the detector x/y pixel grid.

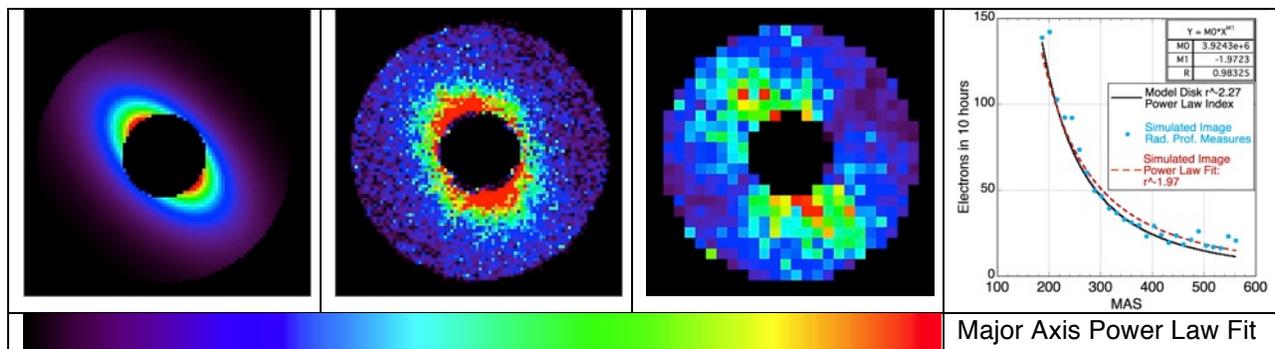

Fig. 7.1. Left Image: Noiseless, 550 nm, 47 UMa 30 zodi disk model at CGI pixel scale, flux density (linear color encoded scale bar): $1 \times 10^{-10}$ Jy to $5 \times 10^{-9}$ Jy pixel$^{-1}$. Middle Image: Simulated SPC image including residual "speckles" and photon noise with 10h integration time, 0 to 115 photons pixel$^{-1}$. Right Image: Per resolution element SNR map: 1 to 15. Graph: Estimation of the disk major axis power law index best fit from its measured SB profile (with noiseless model image normalized to the brightness of the simulated image after PSF convolution).

**Results.** This single image: (a) Detects the disk **with ~ 6 – 15 σ resel$^{-1}$** significance along the disk major axis, that then, (b) closely recovers the disk global morphology and extent, (c) enables a quantitative determination of the major axis radial SB profile as $r^{-1.97}$ (see Fig 7.1 graph) in good agreement with the model image (power-law index ±0.3, invertible to surface density with similar uncertainty), and (d) is devoid of image artifacts that could be mis-interpreted (otherwise) through "breaks" in its radial SB profile as planet-presence or radial differentiation in grain properties.

### 7.2 Multiple "Roll" Observations – a recommended strategy for disk characterization

The residual stellocentric contrast field (i.e., "speckles") is rotationally invariant at the detector image plane with a reorientation of the target scene if the telescope (spacecraft) is rolled about the target axis. To the extent that the residual "speckles" are (at least quasi-) stable with observations at two or more spacecraft "roll" angles, the speckle structure in co-registered target images will stochastically (partially) decorrelate with coaddition (averaging, or medianing with three or more images) in a common celestial frame. This further improves disk detectability, recovery of astrophysical image structure, and overall image fidelity. This approach has been highly successful with spaced-based optical coronagraphic imaging of the outer, EKB, regions of CDSs accessible with *HST* (e.g., Schneider et al. 2014) and is equally applicable at the more demanding combined contrast and IWA delivered by an AFTA CGI.

---

[7] 1 "zodi" of dust has a face on optical depth of $7.1 \times 10^{-8} \, r^{-0.34}$ (where r is the stellocentric distance in AU) and corresponds to the amount of dust in our solar system.

In Fig. 7.2 we illustrate a "three roll" observation of the same 47 Uma hypothetical disk considered in § 7.1 but, in further simulation, observed at three orientation angles w.r.t. the CGI instrument frame differing from the prior 45° major axis position angle also by ±15°[8]. The individual 10 hour simulated exposures are shown at the bottom of Fig. 7.2 with disk major axis PAs = 30° (A), 45° (B), 60° CCW (C) from vertical in the CGI frame. All three images are then differentially rotated (not shown) to a common (image B) celestial frame, wherein the underlying speckle pattern then co-rotates (top panels A – C) with the individual images. With differential rotation, the speckle pattern partially decorrelates, providing a spatially smoother background when coaded (panel D top), as evidenced in the then higher fidelity disk image (panel D, bottom).

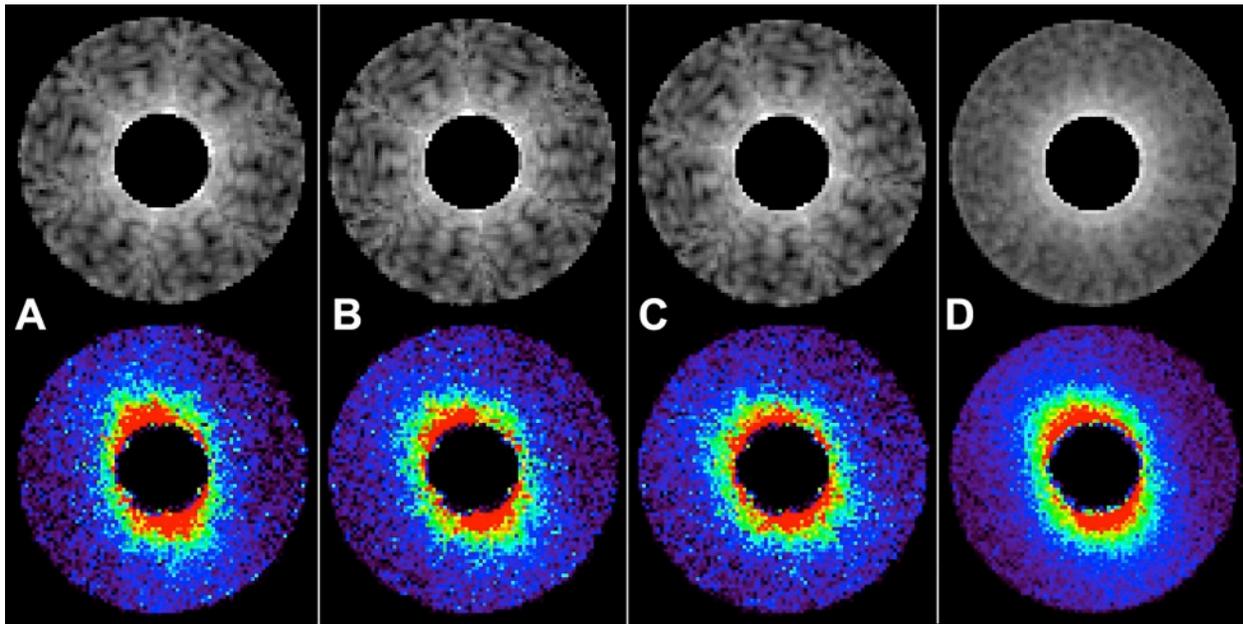

Fig. 7.2. A-C Top: Underlying SPC "56° wedge" mask speckle pattern (synthesized circumferentially over a 360° field by replication) after rotating the disk images A & C by +15°/-15° CCW to a common celestial orientation, and then averaged in panel D. Shown in a $\log_{10}$ stretch (to best illustrate speckle structure) from 1 (black) to 100 photons (white) pixel$^{-1}$ without photon noise for illustrative purposes. A-C Bottom: Individual 10 hour CGI frame simulated exposures, as in Figure 7.1 (for panel B identically) with the disk at 30°, 45°, and 60° CCW (before rotation), then combined (averaged) in panel D after co-rotation to a common celestial frame. Same color stretch as in Figure 7.1.

"N"-roll observations are, of course, N times more expensive of spacecraft time-on-target. E.g., in this case consuming 30 hours (with N=3) to obtain the final image in Fig. 7.2 (panel D). A serendipitous added benefit of this multi-roll observing strategy is also a root(N) improvement in photon SNR, separate from decorrelating the residual speckle structure. N-roll observations should be considered at the outset for high-priority CDS targets, or as a follow-up to improve upon observations of newly discovered zodiacal disk analogs from N=1 survey observations.

## 7.3 An Exoplanetary Zodiacal Debris Ring: Evidence for and Constraining Planet Presence

A second disk simulation is for a nearby star of particular interest, ε Eridani (K2V; 3.4 pc;

---

[8] For illustration we assume a spacecraft off-nominal roll capability of ±15°. Any differential roll capability sufficiently large to rotationally decorrelate image artifact structures on the spatial scale of the "speckles" at the IWA would suffice for this purpose.

V=3.73; age ~ 850 Myr). ε Eri possesses an ~ 35 – 75 AU radius EKB debris ring that is spatially resolved (and "clumpy") at 850 μm. This outer disk, with peak emission at 60 AU (= 18" due to its close proximity to Earth), is (yet) undetected in scattered light, and is estimated to contain ~ 1/6 $m_{lunar}$ of debris dust. Spitzer observations additionally suggest two belts of warm dust orbiting closer to the star as modeled by Backman et al. 2009 at stellocentric distances ~ 20 AU (Δr/D = 0.05) and ~ 3 AU (Δr/D = 0.17; of non-volatile silicate grains as suggested, e.g., by Reidemeister et al. 2011) with optical depth and mass density scaling with Δr/D. Δr/D is the ratio of an annular debris ring's width (Δr = inner to outer edge distance) to its stellocentric diameter. The detected presence (or lack thereof) of such a ring, and "sharpness" of its edge profiles revealed by spatially resolved imaging, puts dynamical constrains on the existence co-orbiting planets, their orbits and masses, and on the locations of parent bodies responsible for the light-scattering dust (Rodigas et al. 2014).

Eps Eri has a radial velocity detected exoplanet candidate with an astrometrically determined mass of 1.55±0.24 $M_{jup}$ and orbital semi-major axis of 3.39 AU (Benedict et al. 2006). Its orbital eccentricity is uncertain, with *e* ranging from 0.25 (Butler et al. 2006) to 0.70 (Benedict et al. *ibid*). The larger eccentricity (in particular) has been suggested as inconsistent with the presence of the Spitzer-inferred innermost zodiacal belt, since a ring-crossing orbit for the posited planet could dynamically disrupt the ring. A second unconfirmed planet, ε Eri c, has been suggested with *a* ~ 40 AU, and Backman et al. 2009 comment that the CDS architecture, suggested from modeling IR to mm wavelength observations, may require three planets.

Here we test the ability of an AFTA CGI to detect and resolve a posited centrally-cleared ring-like zodiacal belt, in some respects "similar" to, but more massive than, our SS's Earth-resonant ring (Dermott et al. 1994) that traps dust particles between 0.8 and 1.3 AU from our Sun. The architecture of a posited ε Eri planetary system is not well established, nor known how additional inner planets (if any) may actually effect an inner zodiacal belt dust distribution. With those caveats, we place such a hypothetical ring mid-way between the suspected 1.55 $M_{jup}$ planet's semi-major axis and the star, with an architecture suggested by and scaled from our SS's main asteroid belt; for this ε Eri model with an inner edge at 1.4 AU, outer edge at 2.28 AU, and maximum dust density at 1.75 AU. We consider total dust masses equivalent to both 30 and 100 zodis and particles with an albedo of 0.17 that scatter isotropically. We assume no other dust components, nor the planet itself, in these image simulations. While an inner planetary system (and hypothetical resonant ring) need not be co-planer with the outer EKB ring that is inclined 65° from face on, that inclination is in good agreement with that of the putative ε Eri b planet (Benedict et al. 2006), and we adopt this inclination also for this zodiacal disk image simulation.

The instrumental configuration (SPC, IWA/OWA, filter band and bandwidth) is identical to the earlier presented "featureless" disk simulation for 47 Uma (§ 7.1). Here too we separately integrate for 10 hours in each of three celestial orientations of the astronomical field off-rolled by -15°, 0° (nominal), and +15°, with "nominal" arbitrarily at a disk major axis orientation 45° w.r.t. the detector "x/y" focal plane image orientation. Results are shown in Fig. 7.3. The debris ring is clearly detected, and well resolved, in each of the three off-rolled images (panels A, B, & C with 30 zodis of dust mass, and with better clarity in F, G, and H with 100 zodis), and its centrally-cleared architecture is unambiguously revealed along with its morphology and geometry. Any one of these single images would serve well in a "discovery mode" survey, and provide a preliminary characterization of the zodiacal ring architecture and photometry. Combining the three images for each case separately into a common celestial frame (panels D or I) improves both the fidelity of the disk image by partial decorrelation of the residual speckle structure (and thus ability to better recover the debris system two-dimensional structure) and enables photometry with peak SNR resel$^{-1}$ at the ring ansae of ~ 15 and 40 (panels E and J, respectively).

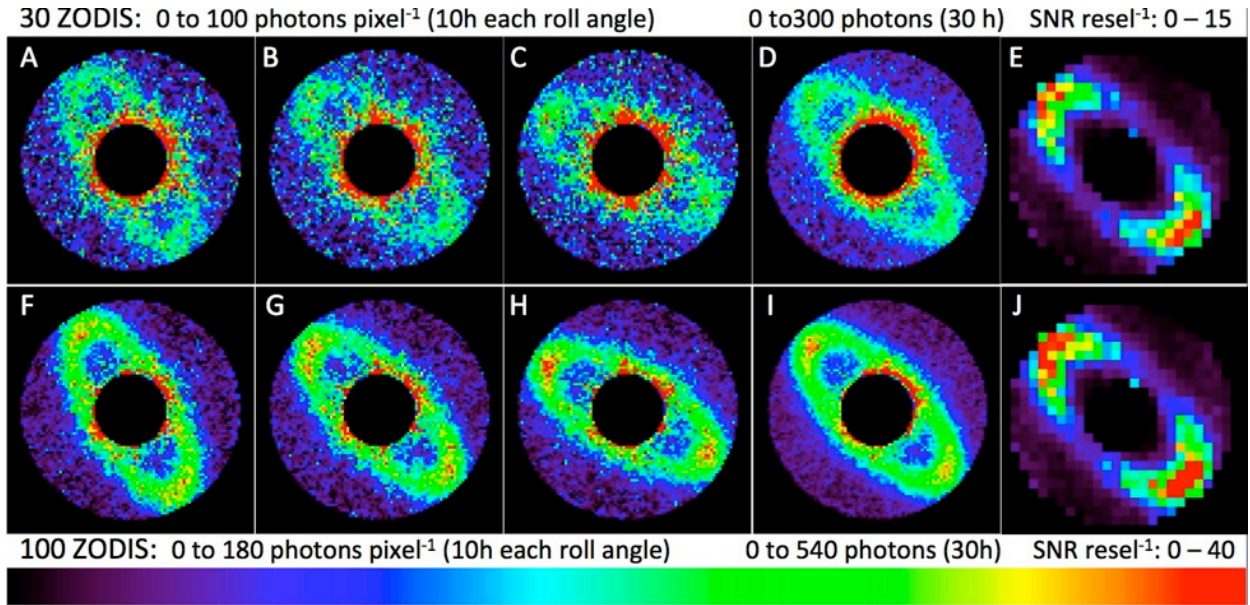

Fig 7.3. Disk image simulations for hypothetical 30 (top) and 100 (bottom) zodi inner debris rings circumscribing ε Eri at r = 1.75 AU. Left three panels (each) differentially oriented on the sky by -15°, 0°, +15°. Panels D & I (each) combine the corresponding three off-rolled images with resulting SNR resel$^{-1}$ maps indicated in panels E & J.

In Fig. 7.4 we compare, for the 100 zodi case, the noiseless debris ring model (top left) to the three-roll combined simulated image (bottom left, identical to Fig. 7.3I, but rotated to put the disk major axis on the image horizontal). A three pixel (averaged) wide, IWA-masked, radial profile cut along the disk major axis centered on the location of the star is shown for the CGI simulated image in comparison to the disk model. Inwardly increasing residual light from the star dominates over the inwardly declining disk SB at r < 375 mas. At larger stellocentric angles, the radial structure of the debris ring is recovered with high fidelity, though broadened at r ≤ $r_{peak}$.

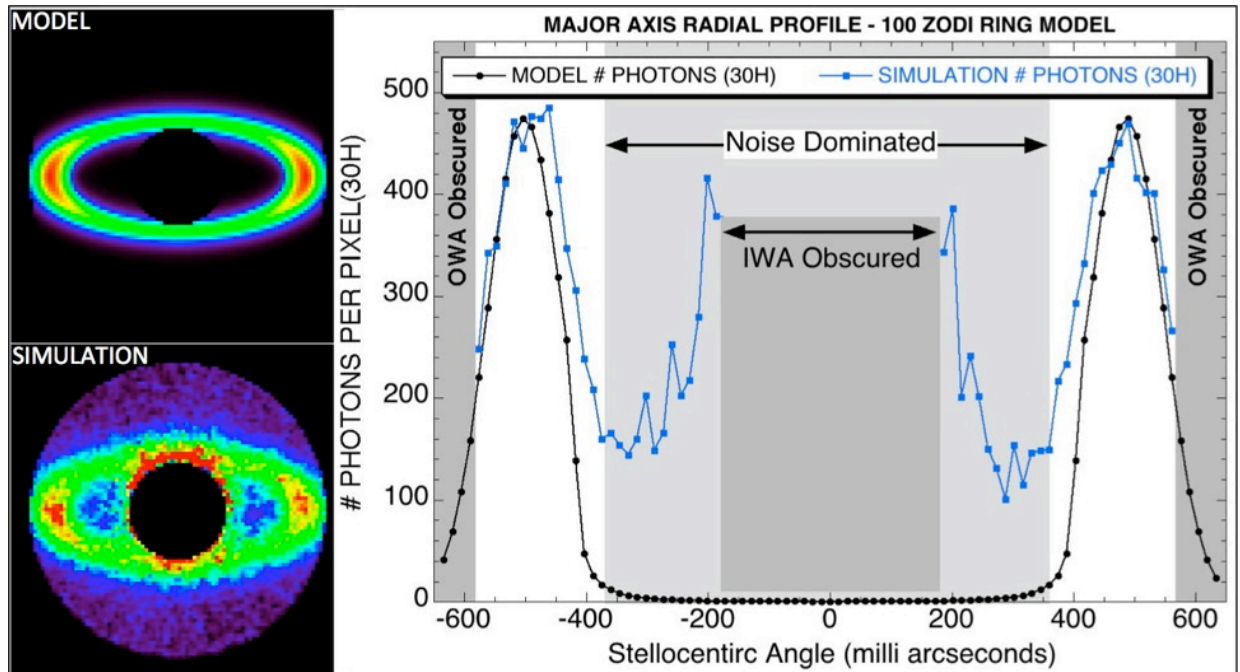

Fig. 7.4. Comparison of noiseless model and CGI simulation of the 100 zodi debris ring scattered-light images with per pixel radial SB profiles along the disk major axis. The OWA of the CGI simulation with the SPC mask obscures the exo-ansal part of the ring at r ≥ 583 mas, beyond the $FWHM_{outer}$ = 569 mas radius of the simulated ring peak SB.

**Results.** In particular, **with SNR ~ 20-40 resel$^{-1}$** (Fig 7.3J) from the half-power points across the ring ansa: (1) the SB peak locations are precisely recovered. (2) Δr is increased by only 4% of D (39.5 mas for this case), primarily as a consequence of the stray-light interior to the ring peaks but also in lesser part in convolution with instrumental PSF (see § 7.5, and relevant discussion vis-à-vis the morphologically similar HR 4796A debris ring by Schneider et al. 1999 and 2009). (3) the ring-edge slopes are closely preserved. (1–3) together provide a high-fidelity tracer of planet presence, orbital chaotic exclusion zone location, and planet mass characterized by the ring semimajor axis: ($a_{HWHM\_IN}$-$a_{HWHM\_OUT}$/$a_{PEAK}$) per Rodigas et al. 2014 (*c.f.*, eq. 1, 2, and 5).

### 7.4. Pericentric Offsets & Non-Isotropic Scattering Phase Functions (SPFs)

**Pericentric Offsets:** The presence of a (massive) planet can perturb otherwise stellocentric dust particle circular orbits, resulting in a pericentrically offset debris ring (Wyatt et al. 1999). This is seen, for example, for Fomalhaut's EKB debris ring offset by ~ 11% of its radius that may be due to the imaged Fom b planet seen just interior to the ring (Kalas et al. 2013). The more massive (~ 10,000 zodi) HR 4796A debris ring has a smaller pericentric offset of ~ 2% of the ring radius (Schneider et al. 2009), though no planet is seen (but is speculatively inferred). A pericentric offset will not only displace a debris ring center, but also will result in diametrically opposed Δr$^{-2}$ SB asymmetries in the disk plane from particles closer to, and further, from the star.

**Non-Isotropic SPFs:** Light scattered by CS grains may have directionally preferential SPFs with dependencies on grain sizes (compared to the observational wavelength) and properties that can be betrayed (or constrained) by azimuthal SB variations in the projected plane of the disk (to the line-of-sight of the observer). I.e., with scattering phase angle (SPA; φ = cos$^{-1}$[sin $i$ cos θ]; with $i$ = inclination and θ = disk-plane azimuth angle.) E.g., see Schneider et al. 2006. SPA SB dependence is unexplored with face-on disks, and is fully explored for edge-on disks, but in the latter case is conflated by the increased optical depth and quadrant degeneracy through the dust along the edge-on line-of-sight. Intermediate inclination disks, though not fully probing the range of SPAs, are the best targets for modeling and interpretation of observationally derived SPFs.

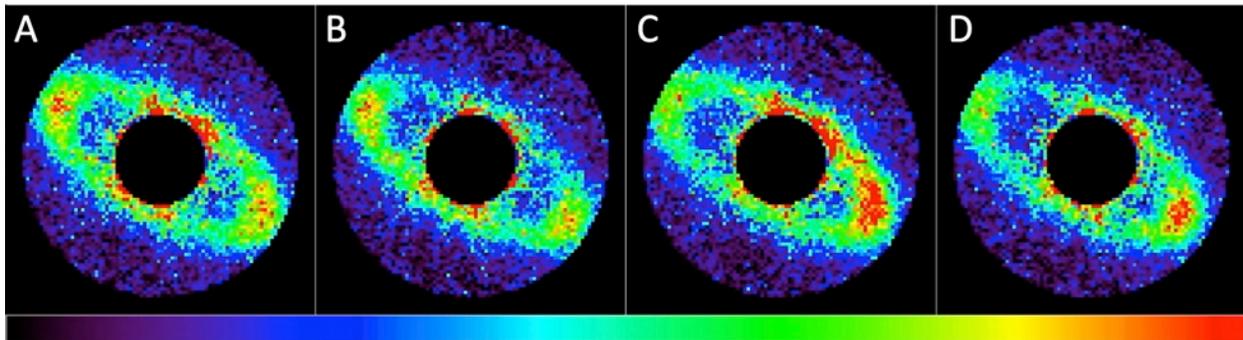

Fig 7.5. AFTA CGI simulated 10h SB images (0 to 180 photons pixel$^{-1}$ linear dynamic range in all panels) of a 100 zodi debris ring per § 7.2 formulation. (A) as shown previously with isotropic scattering, SNR resel$^{-1}$ ~ 15 at the ring ansae. (B) with a non-isotropic SPF per Hong (1985) with a three-component Henyey-Greenstein linear superposition prescription. (C) isotropic scattering but with an 11% $r_{ring}$ stellocentric offset from the central star with pericenter at θ = 335° (CCW from image vertical) in the ring-plane. (D) simultaneously the same non-isotropic scattering as in (B) and the pericentric offset as in (C). The imaged SB asymmetries discriminate between possibilities.

Fig 7.5 illustrates the AFTA CGI sensitivity to both pericentric offsets and non-isotropic SPFs for a hypothetical 100 zodi ε Eri zodiacal-belt per § 7.2 with single 10 h exposures. Panel A is for an isotropically scattering debris ring with no pericentric offset (as shown prior in Fig.

7.2G). In panel B we introduce and recover a Hong 1985 SPF[9], appropriate at λ < 4 μm, with a strong "front to back" asymmetry in scattering efficiency with SPA. Such a strong asymmetry is seen in polarized light (only) for the HR 4796A debris ring (Fig 8.7) with (yet) uncertain interpretation. In panel C, with isotropic scattering, we introduce a disk-plane stellocentric offset of 11% ring radius (scaled from Fomalhaut's EKB ring) arbitrarily with the pericenter 45° CW from the projected ring's "upper" semi-minor axis. Both the $r^{-2}$ pericentric "glow" and the ring's astrometric decentering are recoverable from the image. In panel D we apply both the non-isotropic SPF and the pericentric offset in addition to the still-obvious (OWA-limited) ansal astrometric offsets. This produces, in combination, a morphological intra-ring "gap" more pronounced with bilateral asymmetry on the "back" side of the ring. Interpretation of this image would be better disambiguated with two-band imaging and full-Stokes linear polarimetry (§ 8).

### 7.5. A Note on PSF Convolution (/ Deconvolution)

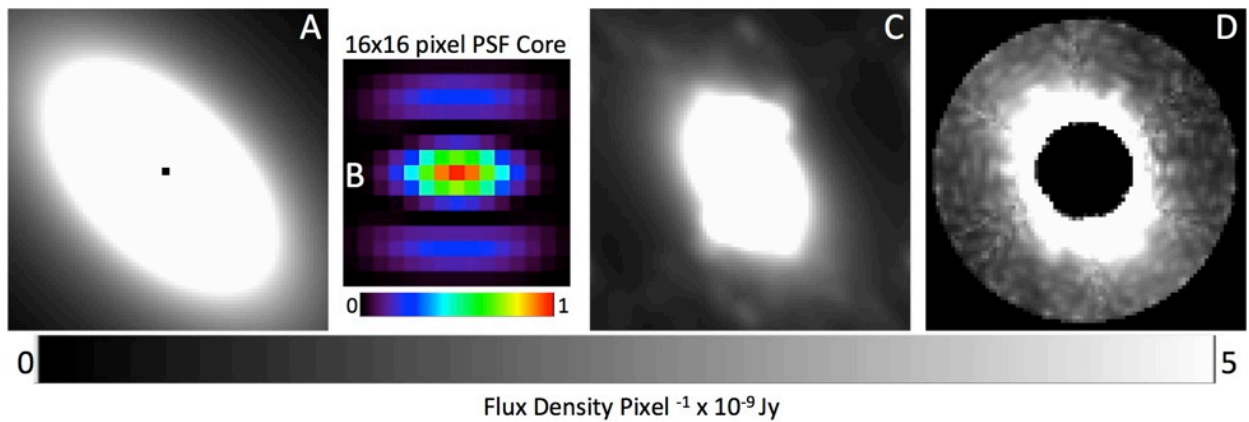

Fig. 7.5. A: Model Disk. B: SPC/wedge-mask PSF core region unity normalized to central pixel brightness. C: Model convolved with the full PSF. D: Convolved model with residual speckles. A & C shown beyond the image-plane mask imposed IWA/OWA.

The complex structure of a spatially variable or asymmetric PSF (e.g., AFTA HLC or SPC, respectively), in convolution with the true, spatially-resolved, astronomical scene can:

(1) broaden spatially resolved features (e.g., narrow or "sharp" edged rings, gaps, clumps and other substructures),
(2) "distort" the two-dimensional morphology of the disk (as in Figs. 7.1 & 7.2), and
(3) result in "flux loss" within the stellocentric working angle limits of the image plane mask.

We comment on each of these below:

(1): The two-dimensional structure of the AFTA HLC PSF is field-point dependent. Thus, to fully recover the detailed morphology of a spatially-resolved source (i.e., CS disk) with high precision may require anamorphic or other field-dependent PSF deconvolution/reconstruction techniques (e.g., Lauer et al. 2002) but, at low SNR, may be difficult if not problematic.

---

[9] The Hong 1985 formulation is a Henyey-Greenstein (HG; 1941) SPF representation using a linear combination of SPA asymmetry parameters, $g_n$, for SS zodiacal dust. We do not suggest this is necessarily the best representation for a physical disk about ε Eri, but use it for demonstration. Recent analysis of scattered-light imaging of the HD 181327 CDS (Stark et al. 2014) with a single HG asymmetry parameter for its debris ring fails to reproduce the observed SB distribution over the full SPA range. In that case, a two component model with $g_1$ = 0.87 and $g_2$ = -0.30 weighted 86% and 13%, respectively, does much better - with possible implications for two discrete populations of dominant particles in the debris birth ring. This is not a big surprise as what we see may be interpreted as spray of material from a recent collision with an ~ 1% $m_{pluto}$ or larger mass asymmetrically ejected from the birth ring and possibly escaping the system.

(2): Convolution "distortion" is the reason for the apparent clockwise rotation (skew) of the disk major axis orientation in the SPC simulated images in Figs. 7.1 & 7.2 compared to the noiseless model in Fig. 7.1. This effect is illustrated in Fig. 7.5 that shows the noiseless morphology of that model disk (with the central star removed) before (panel A), and after (panel C), PSF convolution with the AFTA SPC PSF (panel B showing its central region) and 56° wedge mask. In panel D, the wedge-mask contrast field (speckles with post-processed x10 attenuation) has been added tiling a full 360° in simulation by abutted sector replication. This is the same image as in Fig. 7.1 but without (for illustrative proposes) the addition of photon noise.

This distortion effect could be mitigated with a more symmetrical PSF (e.g., HLC, but note its field dependence), or potentially with differentially reconstructed modeling from multi-roll observations. In the latter case, further study is required to ascertain the degree of calibratability in the general situation of an arbitrary (two dimensional) disk SB distribution.

(3): Additionally, disk flux within the SPC FOV, with convolution, is not fully conserved and therein is underestimated overall. E.g., compare Fig. 7.6 left & right panels shown in an identical linear stretch. Flux from the disk is also redistributed in convolution exterior to the SPC OWA (e.g., Fig. 7.6 top right), and beyond the field stop is unimaged, and hence is irrecoverable even with image-based deconvolution where an exquisite knowledge of the instrumental PSF exists.

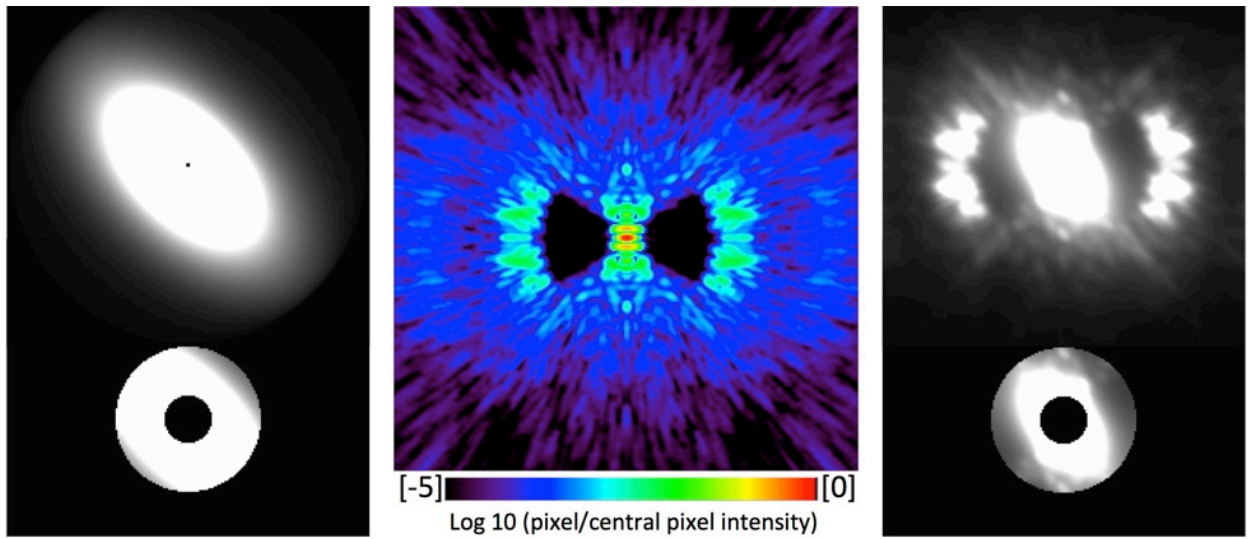

Fig. 7.6. Noiseless disk model (left) in convolution (right) with the "full" (256 x 256 pixel = 3.7" x 3.7") SPC PSF (middle); bottom of each with the SPC field stop. Middle: PSF with $\log_{10}$ stretch normalized to the central pixel brightness. Pixels in green beyond the OWA periphery of the wedge-mask are suppressed by only ~ 0.3% of the central pixel intensity with flux redistributed/enhanced beyond the SPC OWA in the convolved (right panel) image.

## 8. Polarimetry of Circumstellar Debris Systems

Imaging photometry alone, while highly informing in characterizing the distribution and physical properties of CS material, is often not sufficiently constraining to arrive at non-degenerate "answers" to questions such as those posed in § 2. Polarimetry, the additional piece of information that defines the full state of an EM wave scattered from CS material (and planetary atmospheres and surfaces), though often ignored, can provide the key information and insights to arbitrate between conflicting theories in otherwise observationally-ambiguous domains. With wavelength dependence *and* measurement of the polarimetric state, degeneracies between material properties, optical constants, and disk viewing geometries can be broken. For detailed discussion see Perrin et al. 2014a.

## 8.1. Characterizing CS Material With Two-Band Imaging Polarimetry

Debris dust should represent the material being delivered to planet surfaces by asteroids, comets, and meteoroids in (exo-)planetary systems. For an optically thin medium, such as the dust in a CDS, spatially-resolved measurements of the color, the degree of linear polarization (DoP[10]), and the scattering phase angle of light scattered into our line of sight will strongly constrain the grain size, composition and porosity (Fig. 8.1 & 8.2). This, in turn, constrains the type of parent planetesimals (rubble pile or differentiated body) and its volatile (e.g., carbon and ice) content.

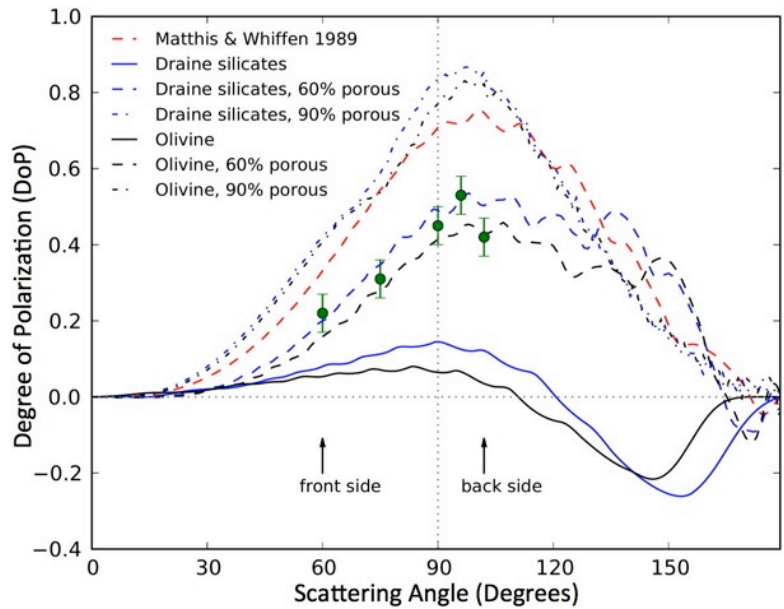

Fig. 8.1. Distinguishing grain properties using spatially-resolved imaging polarimetry: measuring the DoP of dust-scattered starlight as a function of stellocentric azimuthal angle. Shown here, *HST* NICMOS 2 μm coronagraphic polarimetry observations (green points) of the very bright AB Aurigae CS disk, accessible at *HST* contrasts (but IWA = 0.3", in the AFTA CGI working angle range). These spatially resolved polarimetric observations place tight constraints on the likely composition of the light-scattering dust in this transition disk system. In this case 60% porous silicates including olivine that may have once been icy (see § 8.4 discussion on ice lines) provide the best fit. From Perrin et al. 2009.

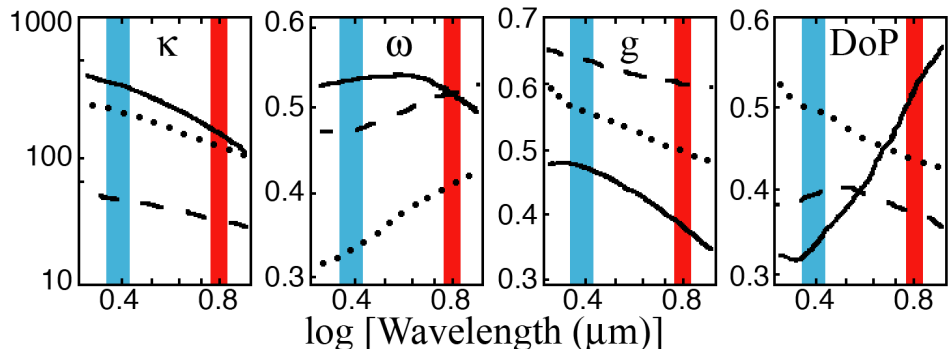

Fig 8.2. CS polarization signals are scattering angle *and* wavelength dependent. Two-band polarimetric imaging observations can measure, and further dis-ambiguate, debris dust properties; here shown over the ~ 2x CGI wavelength range. Disks may be full of volatile-rich porous grains that carry water and carbon to planet surfaces, or may be composed of compact and bone-dry spherules. Different grain types have different wavelength-dependent absorption (**κ**) and scattering (**ω**) efficiencies, directionally preferential scattering phase functions (isotropic: **g**=0, complete forward scattering: g=1 for Henyey-Greenstein scattering), and maximum linear degree of polarization (**DoP**). Examples here compare: (a) compact particles like those in the interstellar medium (solid lines), (b) moderately sized fluffy aggregates (dotted, Cotera et al. 2001) and larger grains inferred in some CS disks (dashed, Wood et al. 2002).

---

[10] The DoP (also called the fractional linear polarization, p, or as a percentage p%) of a polarized source is the polarized intensity normalized by the total intensity (I): $p = (Q^2 + U^2)^{1/2} / I$, derived from the full-linear Stokes pseudovector components. The polarization position angle of the source, $\theta = \frac{1}{2} \tan^{-1} (U/Q)$. Q and U are measured with a set of polarimetric analyzers with the absolute measurement of the total intensity in the same bandpass (e.g., see Hines, Schmidt & Schneider, 2000). With orthogonal-only polarization analysis (such as with o/e ray splitting with a single Wollaston prism) only Q can be measure and p% cannot be determined; see § 8.6).

These properties cannot be unambiguously determined from IR-to-millimeter wavelength spectral energy distributions alone because SED models are degenerate between grain properties and disk radial structure (e.g., see Fig, 8.3). By spatially resolving the disk at two wavelengths *and* in degree of linear polarization (DoP), such degeneracies can be broken (e.g., see Fig 8.4).

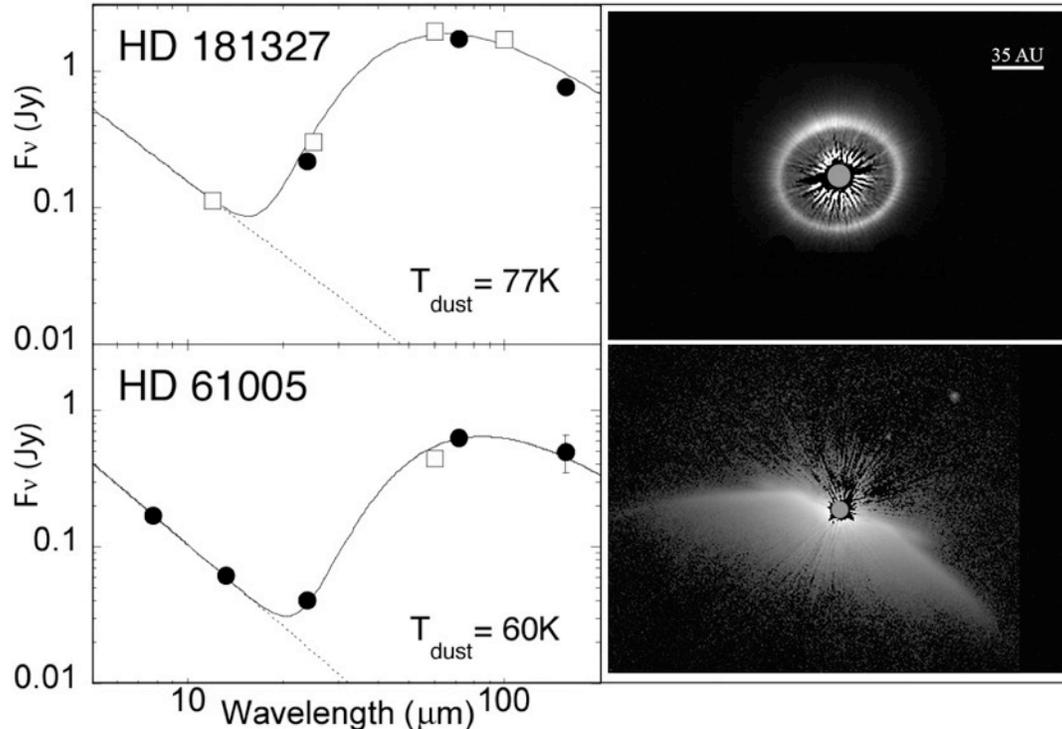

Fig. 8.3. Disk geometries (in particular inclinations) and morphologies cannot be determined by thermal IR excess alone (left). Small particles radiate less efficiently than large particles. Therefore, at a given equilibrium temperature, small particles reside farther from their central stars than large particles. E.g., IR-excesses measured by *IRAS* (squares) and Spitzer (dots) for the dominant particle populations around the CDS hosting stars HD 181327 (top) and HD 61005 (bottom) indicate similar temperatures assuming emission from particles at a single stellocentric distance. Drastically different disk morphologies (and inclinations), however, are revealed (right) with scattered light imaging (*HST*/STIS; Schneider et al. 2014).

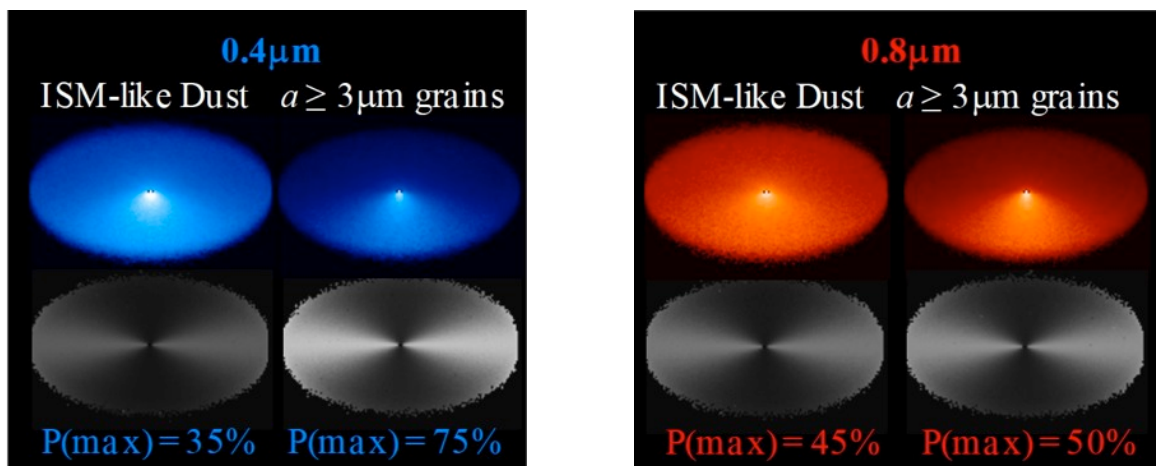

Fig. 8.4. Monte-Carlo scattering models. Top: Total intensity images for CS disks with a surface density $\propto 1/r$. Larger grains show a stronger forward vs. backward scattering asymmetry that is greater at shorter wavelengths. Bottom: Polarization (DoP) maps of the same disks (grayscale; P% = 0% black, P% = 100% white). In both spectral bands, larger grains produce higher linear polarization, but large grains polarize light much more efficiently than smaller grains. Thus, two-band imaging polarimetry provides a powerful diagnostic to constrain grain size distributions.

**Inclinations and Polarization Fraction.** For an optically thin ($\tau < 0.7$) scattering medium, the single-scattering (linear) polarization is a simple function of the scattering angle:

(1) For a small vertical scale height, and a thin ring of scatterers, the polarization is dominated by scattering angles near 90° for edge-on disks (so p% ~ 30 – 40% or possibly higher).
(2) As the inclination increases, the polarization fraction will still be high in regions with projected scattering angles of ~90°.
(3) For a face-on disk, *all* scattering angles are 90° at any stellocentric azimuth angle[11].

In cases (1) and (2) the maximum polarization fraction will be wavelength (as well as particle size) dependent (e.g., see Fig 8.4). For a summary of optical color and grain types over the AFTA CGI wavelength range see § 6.2.

All such grains polarize but (e.g.) porous aggregates, such as grains from icy comets, polarize more than compact grains produced in collisions of differentiated asteroids (Kimura et al. 2006). The DoP, as a function of wavelength, constrains the grain composition and size distribution. The DoP is also a strong function of scattering angle (from up to 100% for right-angle scattering by small grains to zero for forward or backward scattering). Thus, polarimetry is a powerful probe of disk structure including vertical height distribution, inner-hole size, and inclination. For disks that are modestly inclined to the line of sight, the DoP as a function of azimuth angle and color places strong constraints on particle size and porosity as a function of position in the disk (e.g., Perrin et al. 2009; Fig. 8.1). Grains in areas strongly perturbed by planets and showing image structure may have different size distributions than those in quiescent locations.

Once the scattering and polarizing efficiency of disk grains are known, these can be combined with existing longer-wavelength data (e.g., from *IRAS*, Spitzer, Herschel and WISE) to determine the grain absorption efficiency. Together, scattering and absorption efficiency yield the grain albedo, another measure of composition (e.g., see Fig. 5 in Hines et al. 2006). This approach has been successfully demonstrated in modeling *HST* observations of not only of the (*HST*-sensitive) EKB-analog regions of CDSs (HD 181327: Schneider et al. 2006; AU Mic: Graham et al 2007; HD 61005: Maness et al. 2009), but also for protoplanetary disks around young and older classical T Tauri stars (e.g., GM Aurigae: Schneider et al. 2003; IM Lup: Pinte et al. 2008) and transitional disks around Herbig A/F stars (e.g., AB Aurigae: Perrin et al. 2009).

*Identifying the presence of icy and organic-rich grains in disks will give the first clues to the presence of volatiles important for life. Spatially-resolved imaging polarimetry is thus crucial to disentangling the dynamical and compositional history of disks.*

### 8.2. Current State-of-the-Art – Space-Based High-Contrast Polarimetry

DoP maps of the outer (EKB-analog) regions of several CDSs have been produced to date, notably with *HST* polarimetric coronagraphy (a capability that, unfortunately, no longer exists). These include the ~10 My old edge-on disk around the M1 star AU Mic (d = 9.9 pc), and the order of magnitude older, close-solar analog, HD 61005 (G2V; ~100 Myr; d = 35.4 pc pc) that exhibits a complicated dust morphology (Fig. 8.3 and Schneider et al. 2014). Radial changes in polarization at constant stellocentric azimuth angle are indicative of non-homogeneities in the particle size and/or composition distributions. Gradients in these DoP maps inform on the nature of the outer regions of these systems only since the inner (zodiacal-belt analog) regions are beyond current technological capabilities and remain unexplored - but would be in the domain of an AFTA or EXO-S/C CGI.

---

[11] The primary reason for historically-reported decreased polarization in face-on disks with is just "beam dilution" from the projected angular size of the (seeing- or early AO-performance-limited) resel.

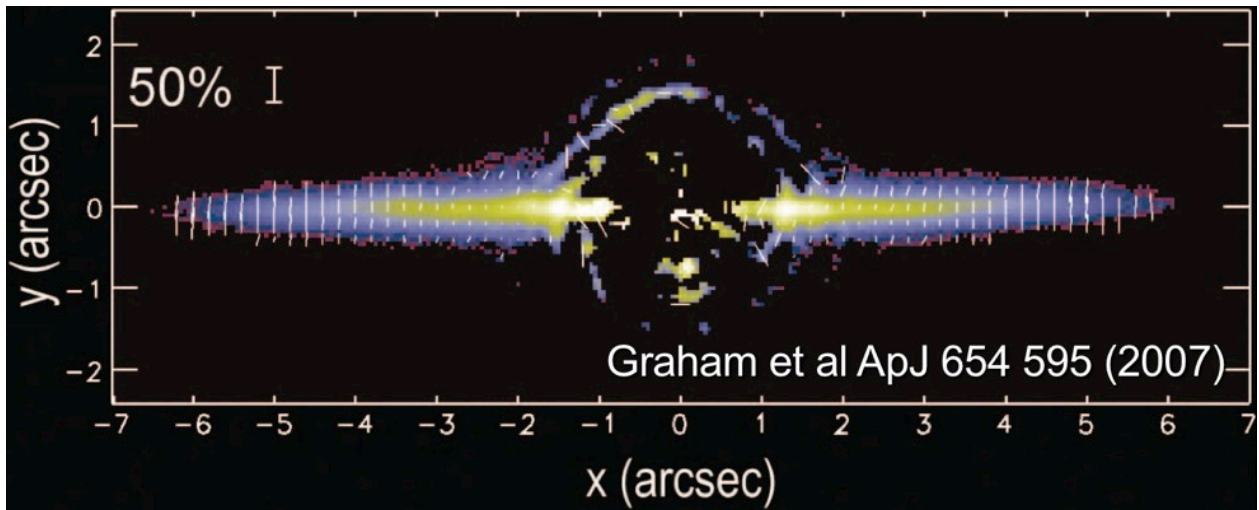

Fig. 8.5. Coronagraphic polarimetry of the AU Mic edge-on CSD system at 0.6 microns (*HST*/ACS) @ r > 2".

In the case of AU Mic (Fig. 8.5; Graham et al. 2007), the polarization at 0.6 μm perpendicular to the disk plane declines from P% ~ 40% at 80 AU stellocentric distance along the disk plane to P% ~ 5% at 20 AU (see Fig. 8.5) indicative of micron size grains of highly porous material (e.g., as found in cometary dust) in an optically thin medium. In combination with disk "colors", these authors model a "hole" interior to an ~ 45 AU collision-dominated birth-ring depleted ~ x300 of infalling grains with ≤ decimeter size parent bodies with evidence for hierarchical grain growth.

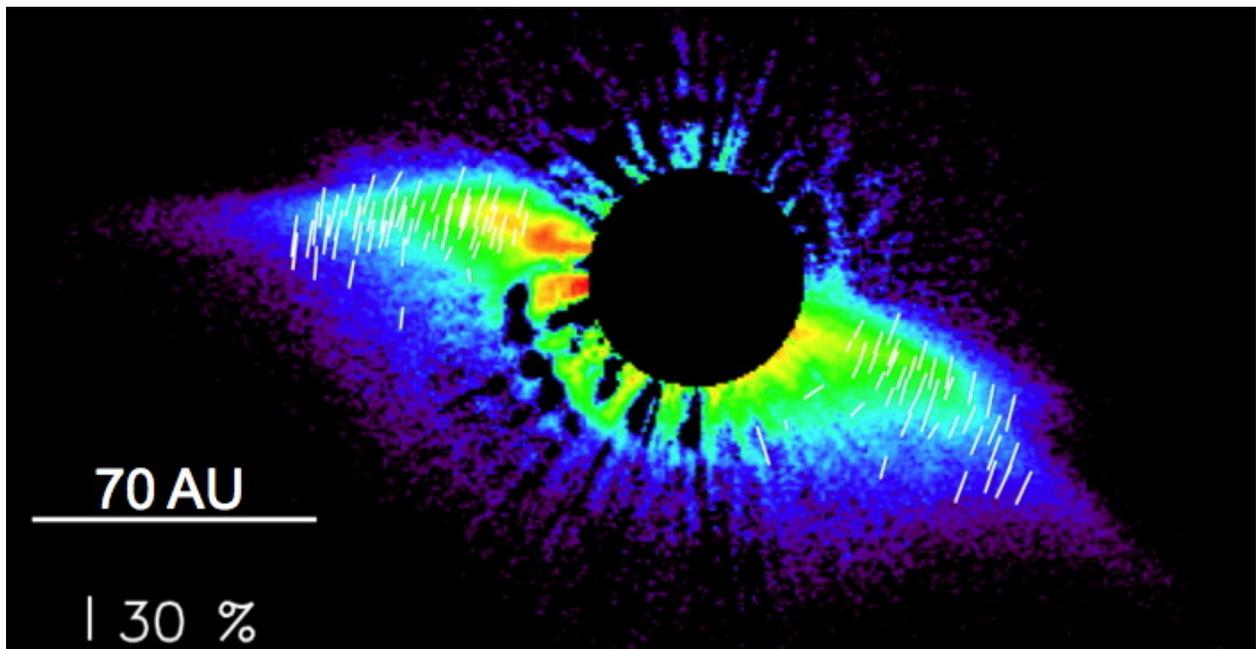

Fig. 8.6. Coronagraphic polarimetry of the HD 61005 ISM-interacting CSD system at 0.6 μm (*HST*/ACS)

In the case of HD 61005 (Fig 8.6), with a high-inclination debris ring and spatially resolved (and at least partially depleted) central "hole" (Schneider et al. 2014), Maness et al. 2009 find (also) an increasing DoP with stellocentric distance in the outer part of this CDS. The complex morphology of the exo-ring CDS has been suggested as possibly arising from a ram pressure interaction as the CDS ploughs through the local ISM (Hines et al. 2007). Small grains may be stripped from the ring and ejected from the system (i.e., mass loss over time). The mapped DoP structure follows (and flanks) the high SB "ridge line", perhaps marking the boundary of a bow shock, with DoP increasing from P% = 15% to 30% at 1.4" < r < 3.6". The polarization signature is also complex and systemically rotates with the swept-back debris structure external to r = 2.1" perpendicular to the (interior) debris ring plane. Mannes et al. 2009 noted that their most promising interstellar gas drag models, discussed therein, reproduced the observed polarization structure.

Both of these systems, as well as the prototypical CDS β Pictoris, exhibit high (tens of percent) DoP. Though the "expectation" for the DoP in their (and other CDSs) currently unobservable zodiacal-belt regions is uncertain, our own solar system's zodiacal dust is highly polarizing (also tens of percent), despite its advanced age. E.gs.: (1) Cometary dust that replenishes the debris in our SS's circumsolar environment can have fractional polarizations as high as ~ 40%. (2) Skylab and Pioneer polarimeters have been used to measure the polarization of zodiacal light as 15-25% seen from 1 AU (Weinberg & Hahn 1980), and (3) polarization of zodiacal light has been estimated to be up to 30% at 1.5 AU at 90° phase angle (Levasseur-Regourd 1996), dropping to ~10% at 40° phase angle at 1.5 AU distance from the Sun. If our own SS's zodiacal dust belt is archetypical in its polarization properties at 4.6 Gyr, similar may be expected for dustier zodi-like disks in exoplanetary debris systems also at advanced ages (?) – a conjecture potentially testable with a CGI, if augmented with a full linear-Stokes polarimeter.

### 8.3. Current State-of-the-Art – Ground-Based High-contrast Polarimetry of CDSs

The new generation of ground-based (EX)/AO instruments measure the polarized intensity (p*i) of circumstellar material, but without (yet) the ability to unambiguously decouple the polarization fraction from the total intensity in high-contrast imaging. This can conflate (and potentially mislead) interpretations of disk structure and properties. E.g., in the very recent (and puzzling) case of the HR 4796A debris ring as imaged by GPI (Perrin et al. 2014b), a K-band "total light" image (Fig 8.7 - left) has a relatively isotropic scattering phase function indicative of small particle Rayleigh scattering. This is in approximate agreement with an *HST*/STIS image at optical wavelengths (Schneider et al. 2009). However, this (Rayleigh scattering) is counter-indicated by the GPI polarized light image as it does not peak at the ring ansae and has a very strong (> x9) minor-axis opposed west-side brighter polarized light asymmetry. The GPI authors speculate that the debris disk (counter to prior interpretation) may be optically thick, or self-shadowed on the east side. While this is a possibility, a P% (DoP) image could arbitrate. Unfortunately, an unbiased DoP (P%) image cannot be produced using the GPI "total light" image for flux calibration. In order to achieve high-contrast in total light, angular differential imaging (ADI) and KLIP (a form of principle value decomposition) post-processing is necessarily employed, but does not preserve the flux density in the ring. The authors estimate as much as 75% of the disk flux density maybe "lost", in a spatially variable manner, not amenable to high precision photometric calibration. p*i imaging provides contrast augmentation, but can lead to ambiguities in interpretation (also see Perrin et al. 2009).

The spatial OWA limits of today's highest-performance and most aggressive EX/AO polarimeters (GPI and SPHERE) are roughly comparable to an AFTA CGI on the red side of its optical wavelength range. Even though they do not have the IWA/contrast reach of an AFTA CGI, as nIR rather than optical imagers, they can provide some symbiotic measures or constraints for wavelength dependent solutions for bright dust in CDSs within their grasp.

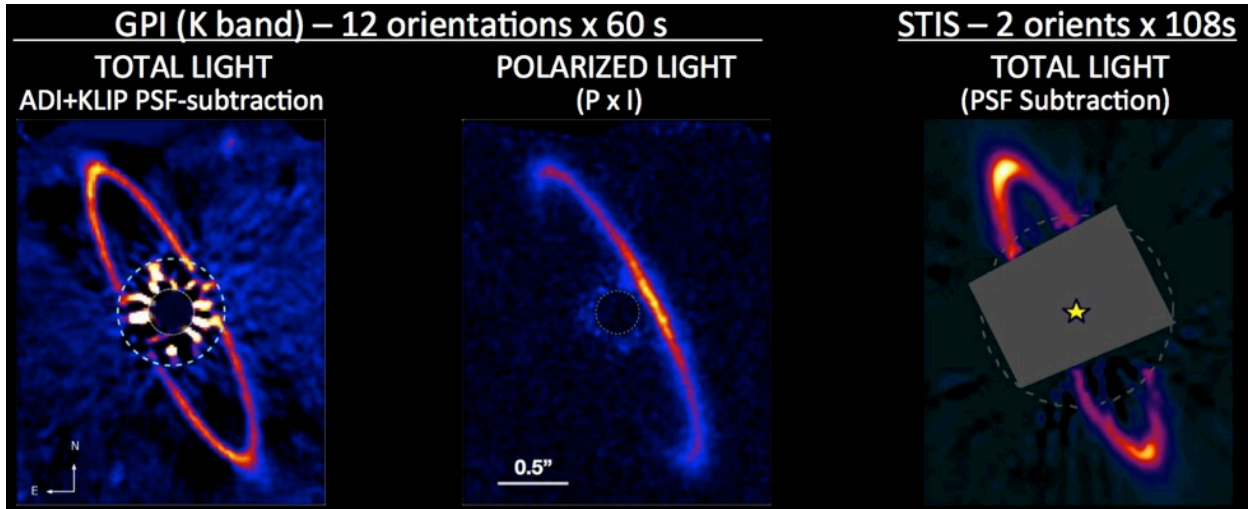

Fig. 8.7. Images of the HR 4796A debris ring. Left and Middle: GPI uncalibrated total light and polarized intensity (p*i) in K-band (Perrin et al. 2014b). Right *HST*/STIS calibrated total light in broadband optical (Schneider et al. 2009).

### 8.4. Zodiacal Disks: Exploiting the AFTA CGI Uniqueness Space with CDS Polarimetry

In § 8.1 we discussed how the physical properties of CS materials may be characterized with imaging polarimetry. These methods are equally valid and applicable to both the outer (EKB-analog) "cold dust" regions of CDS where they have been applied (as exampled in § 8.1) and, contiguously, into the inner (zodiacal-belt analog) "warm dust" regions akin to our own solar system's asteroid belt and habitable zone yet unexplored in exoplanetary debris systems. AFTA CGI post-processed image contrasts ~ $10^{-9}$ resel$^{-1}$ in the subarcsecond domain will enable exploration of these warm dust regions at optical wavelengths in exoplanetary systems with architectures similar to our SS's potentially possessing as little as 10x the dust in our SS's zodiacal dust belt (see Fig. 2.2). Elucidating and constraining the material composition, particle geometry, size distributions, and degree of structural complexity (e.g. porosity) of CS materials in combination, augmented with CGI-enabled spatially-resolved polarimetry, is highly compelling. Perhaps more so in the warm dust region uniquely explorable at CGI contrasts then in the fossilized relics that are the cold dust disks explored to date with (primarily) *HST*.

It is in the warm dust regions where terrestrial planets are posited to form, and where the systemic ice-lines may be found (interior to which mantled volatiles otherwise found in the outer disks have sublimated off of their CS grains). One of the key conundrums in forming terrestrial planets in these "dry" zones is then making them "wet". Investigating CS dust properties in zodiacal disks directly probes this environment, and spatially-resolved imaging polarimetry can inform on the existence of ice lines in zodiacal disks through tell-tail signatures of radial discontinuities in CS grain polarizations. The dynamical history of the terrestrial planet formation in our own SS is being deciphered by studying the composition of the main-belt asteroids and Centaurs, with one of the key puzzles – how did Mars form (e.g., see Walsh et al. 2011) in the context of the "Grand Tack" model[12] of the early evolution of the inner SS? We cannot measure asteroids and Centaur-like objects in exo-solar debris disks, but we can ascertain the dust particle properties from debris liberated by the collision of these parent objects with polarimetric sensitivity to low surface brightness dust uniquely provided with an AFTA CGI.

---

[12] http://www.boulder.swri.edu/~kwalsh/GrandTack.html

## 8.5. AFTA CGI Polarimetric Analysis - Baseline

The baseline CGI technical design incorporates the use of a single, fixed, linear polarimetric analyzer to mitigate the deleterious effects from intrinsic intra-pupil polarization mixing due the variation in beam angle from the primary mirror. In (current) implementation this is done by using only one of two orthogonal polarizations produced by a Wollaston prism for the combined instrumental-plus-astrophysical signal of highest contrast. This may be required to reach a focal-plane image contrast (target-uncorrelated starlight-suppression) requirement, but could be deleterious, rather than beneficial, for science with the realization that both exoplanetary atmospheres and circumstellar materials are intrinsically (and often highly) polarized, as opposed to the troublesome stellar light to be eradicated that is usually essentially unpolarized (p% $\cong$ 0).

With a measure (image) informing on only the polarization component intensity of the signal from a polarized source (CDS or exoplanet) in only one polarimetric-analyzer orientation, the total intensity signal cannot be ascertained. Consider the "extreme" (but likely non-physical) case of a 100% linearly polarized source. If, by happenstance of likely unknown astrophysical systemic geometries, the spacecraft (and image plane fixed to an instrumental polarimetic analysis angle) is oriented such that the analyzer is oriented orthogonal to the polarization pseudovector of the source (planet or dust structure), then no (zero) signal will be measured. In that case disk structures, and planets, would go undetected that otherwise would be visible in "total light" (if sufficiently bright and not dominated by upstream instrumental polarization). This is a loosing scenario.

At the other extreme, with the polarimetric analyzer by chance oriented along the polarization pseudovector of the source, the full total intensity of the source would be measured. But, without knowledge of the orientation of the polarimetric state pseudovector, it is impossible to know if the total flux is measured or only some fraction of it. I.e., with respect to any celestial target's polarization state pseudovector orientation, the instrumentally-fixed analyzer orientation on the sky will be random (unless for some operational reason constrained, but still unknown). Thus, source intensity measures will stochastically (but unknowably) be somewhere between zero and maximum, and will almost always dilute the total intensity signal. In many cases this could make detection problematic, and all photometry impossible or ambiguous at best.

Astrophysical linear polarization fractions (DoP, or p%) of CDS material and exoplanet atmospheres are not expected to be 100%, but are often likely to be on the order of tens of percent (e.g., Stam et al. 2004, Perrin et al. 2014a). For disks (unlike planets as point sources) both the polarization pseudovector component total intensity (I) and the source EM field orientation ($\theta$) will be spatially variable (and resolvable) with stellocentric location in the debris system. This further conflates the photometric issue, and can make mapping even detectable disk structures also problematic. For exoplanets, the intrinsic polarization signal (except in the case of a face-on orbit) will depend on orbit phase with scattering phase angle. This is separate from, but then conflated with, intensity "light curve" change due to the area of the planet illuminated as it orbits. With a single polarimetric analyzer conflating the two combined signals, exoplanet target "revisits" (assuming successful initial identification) - to sample at different orbit phase - would be hampered in interpretation with this degeneracy. Apart from astrophysical interpretation (see below) of material properties, this degeneracy may also be an important (limiting) consideration for differentiating exoplanets from "clumps" in disk structures.

## 8.6. CGI Polarimetric Analysis - Beyond the Baseline

We discuss the science benefits of a CGI polarimetric analysis capability, in particular under the presumption of a polarimetric measurement capability beyond the baseline with full-linear Stokes polarimetry, in § 8.2 (and uniquely for the warm, inner zodiacal, dust regions of CDSs in

§ 8.4). Here we differentiate between the measurement capabilities and the utility of linear Stokes polarmetric systems implementing only a single (current CGI baseline) vs. 2 orthogonal vs. 3 or 4 ("full") phase-angle tiling polarimetric analyzers (see Fig. 8.8).

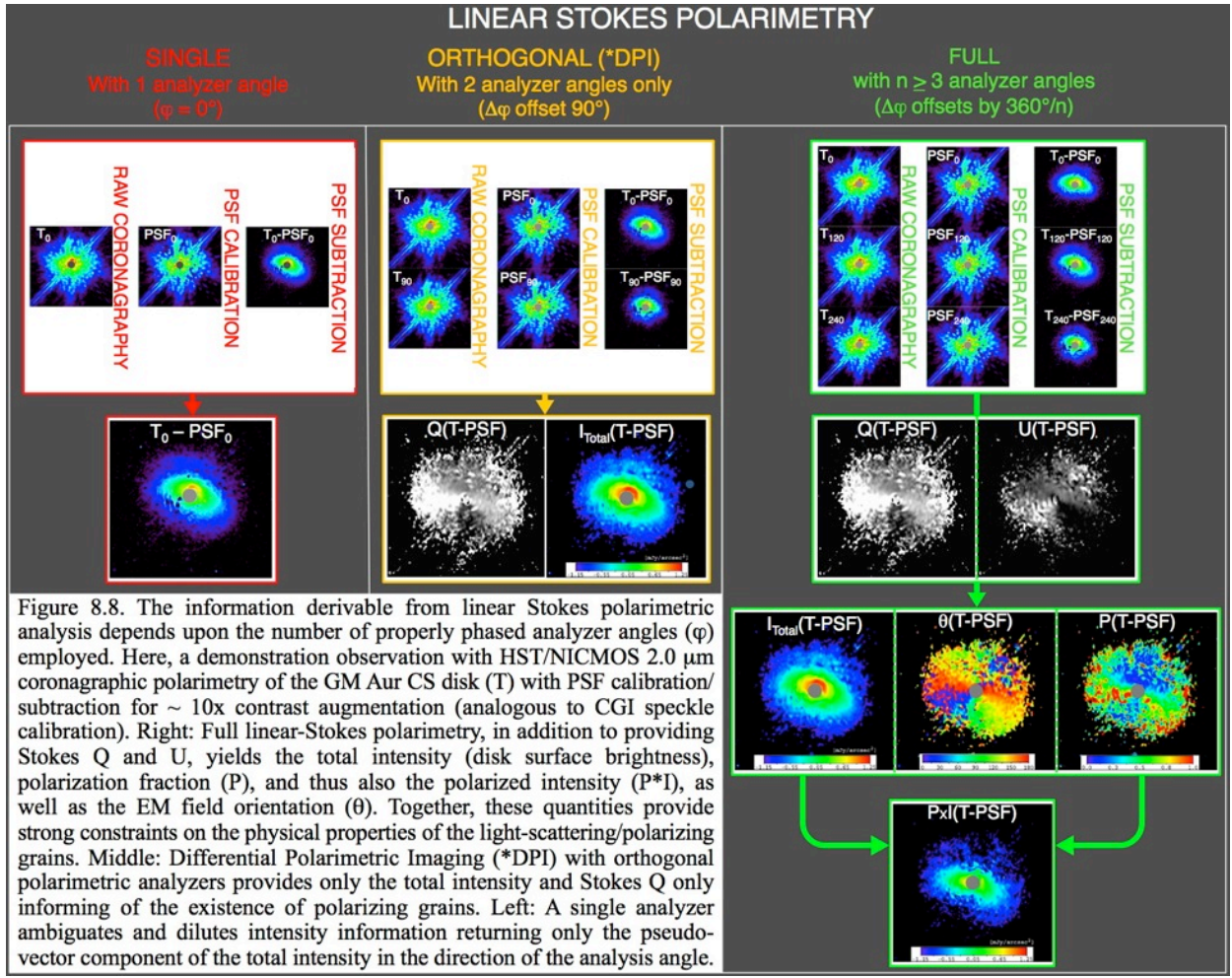

Figure 8.8. The information derivable from linear Stokes polarimetric analysis depends upon the number of properly phased analyzer angles (φ) employed. Here, a demonstration observation with HST/NICMOS 2.0 μm coronagraphic polarimetry of the GM Aur CS disk (T) with PSF calibration/subtraction for ~ 10x contrast augmentation (analogous to CGI speckle calibration). Right: Full linear-Stokes polarimetry, in addition to providing Stokes Q and U, yields the total intensity (disk surface brightness), polarization fraction (P), and thus also the polarized intensity (P*I), as well as the EM field orientation (θ). Together, these quantities provide strong constraints on the physical properties of the light-scattering/polarizing grains. Middle: Differential Polarimetric Imaging (*DPI) with orthogonal polarimetric analyzers provides only the total intensity and Stokes Q only informing of the existence of polarizing grains. Left: A single analyzer ambiguates and dilutes intensity information returning only the pseudo-vector component of the total intensity in the direction of the analysis angle.

1. With only a single linear polarization angle measurement (the CGI baseline) neither the source total intensity, nor its polarization fraction (or scattering-surface EM field orientation) can be measured. *This is not an astrophysically useful mode for diagnostics.* Moreover, by randomly attenuating the (polarized source) signal with an arbitrarily partially-crossed polarimetric analyzer somewhere between 0° and 90°, detectability is hampered and analysis is problematic. There would be an engineering "contrast trade" between using a single polarizer for the purpose of dealing with (reducing the impact of) instrumental polarization and astrophysical signal dilution for a given target's DoP. In different domains this may be a net positive or a net negative outcome. To evaluate those trades a detailed model of the instrumental polarization would be needed.

2. With two (but only two), orthogonal, linear polarization states ("o" and "e") measured, the total intensity signal can be recovered simply as the sum of those two images. This is the same, effectively, as observing in the same bandpass without polarimetric analyzers. The advantage over non-polarimetric images is that the difference signal, Stokes Q, is obtained and will inform on the presence of a polarizing source (useful for disk or planet detection). But (unlike with 3 or 4 polarimetric analysis angles) without Stokes U neither p% nor θ can be ascertained. This is a useful mode over simple imaging with zero polarizers by providing some additional capability. The Stokes Q signal indicates the presence of a polarized source (e.g., differentiating from an ill-placed faint background star), and the problem with one polarizer angle not allowing the recovery of the total intensity signal is mitigated. *This "Differential Polarimetric Imaging" mode has often been employed in ground-based polarimeters.*

3. With three (0°, 120°, 240°) or four (0°, 45°, 90°, 135°) analyzer angles[13,14] ($\varphi$), a full-linear Stokes solution is derivable from the set of (3 or 4) polarimetrically analyzed images providing (Q, U) → I (total intensity), DoP (a.k.a. p or p%) and $\theta$ (the polarization position angle). *These are the diagnostics that are key to constraining compositional and structural analysis in CDS materials and exoplanet atmospheres* (see § 8.1 of this report, and discussion in "Quick Study" report by A. Burrows), in concert with wavelength dependence.

### 8.7. The CGI Full-Linear Stokes Imaging Polarimetry Opportunity

High-contrast (coronagraphic) full-linear Stokes imaging polarimetry is a capability that is not currently planned for any future NASA astrophysics missions, but could be considered for an AFTA or EXO-C/S CGI as a potentially low-cost value-added capability. A CGI full linear-stokes polarimeter, integral to a starlight suppression system with wavefront error control to deliver high contrast, would enable new observational domains into the unexplored zodiacal-belt analog regions of CDSs advancing the state of our knowledge of exoplanetary system formation/evolution mechanisms, properties, and architectures – without ambiguities inherent in p*i or PDI (polarimetric differential imaging).

### 8.8. Notes on Full-Linear Stokes Polarimetry Implementation Architectures

Several different implementation strategies (at least) may be considered to enable full-linear Stokes polarimetry. For the four analyzer angle approach using Wollaston prisms that split the input beam (after the coronagraph) into its orthogonal (o/e) polarizations as follows:

(a) using two Wollaston prisms (W1 and W2), each in turn (inserted into the high contrast beam) producing images in orthogonal (o, e) polarizations imaged simultaneously "side by side" on the detector. The second prism is rotated 45° with respect to the first, so together two pairs of polarimetrically analyzed images are produced as follows: W1(0°, 90°) and W2(45°, 135°). This will allow (for greater simplicity) non-multiplex use of the two prisms, i.e., sequentially imaging the W1(0°, 90°) and W2(45°, 135°) image pairs on the detector (at different locations); this is the EXCEDE concept approach (e.g., Guyon et al. 2012.)
(b) with equal-intensity beam-splitting optics ahead of W1/W2 (which can present a challenge for induced instrumental polarization) all four images could be simultaneously acquired.
(c) alternative to (b) using a single Wollaston prism, with an rotatable half-wave plate (mechanically, 0° and 22.5° rotations) ahead of the prism for sequential imaging in pairs. In pairs: W1(0°, 90°) and (45-degree phase retarded[15]) W1(45°, 135°), sequentially.

---

[13] $\varphi$ is the polarimetric analyzer angle with respect to an image reference plane tied to the detector. $\theta$ is the source polarization position angle in that same reference frame. The celestial PA of the source polarization position angle is a simple rotation of $\theta$ into the sky frame with north PA = 0° and measured CCW from north.
[14] With four such analyzer angles Stokes V (used to measure circular polarization) is additionally derivable, but is unnecessary, since light scattered by CS debris dust will not have a circular polarization signature.
[15] Rotating a half-wave plate through an angle $\alpha$ will rotate the EM field exiting the half-wave plate by 2 x $\alpha$.

In any of these cases, depending upon the spectral channel bandwidth (and residual chromatic error tolerance at the final focal plane), the Wollaston prism design would likely need to be achromatized (for image quality in the final focal plane).

The three analyzer angle approach uses a set of three linear polarizers differentially rotated by 120° to produce a set of three polarimetrically analyzed images: $I_{0°}$, $I_{120°}$, $I_{240°}$.

(a) the simplest approach, often used in space applications with (relative) PSF-stability is for sequential measures in a non-multiplex optical path, selecting each of the three linear polarizers in turn (e.g., on a mechanism wheel). This was the successful approach to *HST* coronagraphic polarimetry.
(b) a variant of (a) is to use a single fixed linear polarizer with a rotating wave plate.
(c) for simultaneous imaging, the beam after the coronagraph may be equal-intensity split into *three* channels with care not to induce instrumental polarizations in the beam splitting (this could be a challenge) ahead of the three fixed (differentially rotated) linear polarizers.

This above describes some conceptual configurations for full-linear Stokes polarimeters in a single spectral channel. Two spectral channels can be accommodated either without or with spectral multiplexing (e.g., dichroic red/blue beam splitting ahead of any of the configurations described), with then twice as many achromatized Wollaston prisms or linear polarizers. Spectrally splitting after a single Wollaston prism (rather than before two Wollastons prisms) is likely non-viable as achromatizing a single prism over an ~ 2x wavelength range could pose an insurmountable image quality challenge.

One can also achieve full linear-Stokes polarimetric analysis with a single polarizing element with an analyzer angle fixed in the frame of the detector by "rolling" the telescope to change the image orientation formed on the detector (e.g., sufficiency with three rolls with 0°, 60°, 120° differential orientations). This is something that should be considered if in the long run a polarimetry capability more suitable for scientific purposes cannot be accommodated with a CGI (though rotational distortion due to convolution with an asymmetric PSF can complicate calibration). This may not be possible for some (or any) target pointings at a single epoch due to spacecraft pointing constraints. If so, it may be possible to revisit targets at three separate epochs (e.g., months apart) to allow the "sky" to re-orient w.r.t. spacecraft nominal roll over time. One consideration in this latter scenario is that the revisit timescale must be shorter than the Keplerian timescale for "significant" motion either an exoplanet or the smallest debris dust clumps resolvable, since the target polarization fraction is scattering phase angle dependent.

### 8.9. Image Contrast Augmentation with "Polarimetric PSF Nulling" (PPN)

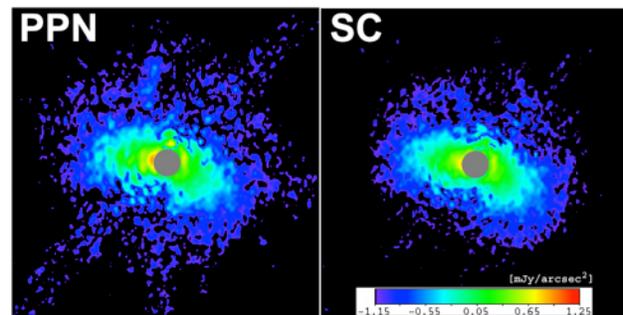

Fig. 8.9. Polarized Intensity (p*i) imaging of the GM Aurigae CS disk from full-linear Stokes (HST/NICMOS 2.0 micron) polarimetry with PPN only (left) and with speckle (PSF template) calibration (SC) and subtraction

Full-Linear Stokes imaging polarimetry, that enables measurement of the polarized intensity (p*i) of polarizing light-scattering material, *also* importantly can provide order-of-magnitude (or larger) contrast augmentation when imaging polarizing circumstellar material (typically ~ tens of percent polarization for CS debris and many exoplanetary surfaces). With full-linear Stokes imaging, unpolarized coronagraphic residuals (e.g., "speckles") that are correlated with the stellar light from the (largely) unpolarized ($p_{star}$ = 0) PSF halo can be suppressed, in principle, to the limit of the photon noise for a fully unpolarized star in p*i imaging (where p = $p_{star}$ + $p_{disk}$). This PPN method was successfully demonstrated in space with *HST* coronagraphic polarimetry (e.g., Fig. 8.9, at less aggressive raw contrasts than an AFTA CGI will deliver) *and works independent of, and multiplicatively with, other starlight/speckle suppression methods*.

(right). The (input) set of raw coronagraphic polarization analysis images ($I_{0°}$, $I_{120°}$, $I_{240°}$), as shown in Fig. 8.8 (top, right), are dominated by residual unpolarized starlight in the stellar halo after incomplete coronagraphic suppression. PPN *without* post-processing/speckle calibration (above, left) removes most of this residual stellar light revealing the underlying polarized-light disk structure in a manner nearly comparable to (but without) SC (above, right). Post-processing methods such as SC envisioned for AFTA CGI may be combined with PPN for multiplicative contrast augmentation in polarized intensity imaging (to the photon noise limit).

## 8.10. A Practical Note on Signal Requirements for Full Linear-Stokes Imaging Polarimetry

While the linear Stokes parameters Q and U (or as normalized by the total intensity, I, such that q = Q/I and u = U/I) follow Poisson statistics, the derived polarization fraction (e.g., see Hines, Schneider and Schmidt 2000) is a positive definite quantity, and follows a Rice distribution function:

$$f(x \mid \nu, \sigma) = \frac{x}{\sigma^2} \exp\left(\frac{-(x^2 + \nu^2)}{2\sigma^2}\right) I_0\left(\frac{x\nu}{\sigma^2}\right),$$

where $I_0$ is a zero order modified Bessel function of the first kind.

Thus, the requirement on the per resolution element uncertainty in the polarization fraction, p, is such that $p/\sigma_p \geq 4$. For a 50% linearly polarized source $\sigma_u \sim \sigma_q \sim \sigma_p$ and requisite integration times for the polarimetric analysis "input" images (e.g.: [$I_{0°}/I_{90°}$], [$I_{45°}/I_{135°}$] from two Wollaston prism polarimetric analyzers) must be exposed to obtain $p/\sigma_p \geq 4$. E.g., in approximation, for a 12% polarized source, ~ 99% confidence in polarization fraction measures requires $\sigma_p$ < 3%. For full treatments of polarization uncertainties see Simmons & Steward (1985) and Montier et al. (2014).

## Acknowledgements

The author is most thankful for the collaborations, helpful suggestions, and information provided by Dean Hines on issues of polarimetry, Thomas Greene on AFTA CGI image simulations, Mark Kuchner and Christopher Stark on scattered light models and use of their Zodipic code, Wesley Traub for details of anticipated WFIRST/AFTA performance expectations and limitations, John Krist for PSF and coronagraphic reference data, and the members of HST/GO 12228 team in the development of the science case discussed herein. He also wishes to thank Aki Roberge, James Graham, Holly Maness, and Marshall Perrin for permissions to reproduce their illustrative figures and their insightful discussions related to them. The author gratefully acknowledge NASA's Exoplanet Exploration Program for supporting this study on behalf of the WFIRST/AFTA Science Definition Team and the Exo-S and Exo-C Science and Technology Definition Teams.

## References

Aumann, H. H., Beichman, C. A., Gillett, F., et al. 1984, ApJ, 278, 23
Backman, D., Marengo, M., Stapelfeldt, K., et al. 2009, AJ, 690, 1522
Benedict, G., McArthur, B., Gatewood, G., et al. 2006, AJ, 132, 2206
Boley, A. C., Payne, M. J., Corder, S., et al. 2012, ApJ, 750, 21
Butler, R. P., Wright, J. Y., Marcy, G. W., et al. 2006, ApJ, 646, 505
Cotera, A., Whitney, B., A., Young, E., et al. 2001, ApJ, 556, 958
Debes, J., Weinberger, A. J., Schneider, G. 2008, ApJ, 673, 191
Maness, H., Kalas, P., Peek, K. M. G. 2009, ApJ, 707, 1098
Montier, L., Plaszczynski, S., Levrier, F., et al. 2014, arXiv:1407.0178
Perrin, M. D., et al. 2014b, ApJ, submitted; arXiv:1407.2495.pdf
Perrin, M. D., Hines, D. C., Wisniewski, J. P., Schneider, G. 2014a, in Polarization of Stars and Planetary Systems, in press.
Perrin, M. D., Schneider, G., Duchene, G., et al. 2009, ApJ, 707, L132
Pinte, C. 2008, 489, 633

Decin, G., Dominik, C., Waters, et al., C. 2003, ApJ, 598, 636
Dermot, S. F., Stanley, F., Jayaraman, S. 1994, Nature, 109, 241
ESO14, 2014, PR# eso1436; http://www.eso.org/public/news/eso1436
Graham, J. R, Kalas, P. G., Matthews, B. C. 2007, ApJ, 654, 595
Guyon, O., Schneider, G., Belikov, R., et al. 2012, SPIE, 8842, 1
Henyey, L. G., & Greenstein, J. L. 1941, ApJ, 93, 70
Hines, D. C. 2007, ApJ, 671, 165
Hines, D. C., Backman, D., Boumann, J., et al. 2006, ApJ, 638, 1070
Hines, D. C., Schmidt, G., Schneider, G. 2000, PASP, 112, 983
Hong, S. S., 1985, A&A, 146, 69
Kalas, P., Graham, J., Fitzgerald, M. et al. 2013, ApJ, 775, 56
Kant, I. 1755, in Zeitz, Bei W. Webel, 1798. Neue aufl.
Kelsall, T., Weiland, J., Franz, B., et al. 1998, ApJ, 508, 43
Kimura, H., Kolokolova, L., Mann, I. 2006, A&A, 449, 1243
Laplace, P.-S. 1796, in Exposition du systeme du monde
Lauer, T. 2002, SPIE, 4847, 167
Levasseur-Regourd, A. C. 1996, IAU Colloq. 150104, 301
Reidemeister, M., Krivov, A., Stark, C., et al. 2011, A&A, 527, A57
Rodigas, T. J., Malhorta, R., and Hinz, P. M. 2013, ApJ, 780, 65
Saffe, D., Domez, M., and Chavero, C. 2005, A&A, 443, 609
Schneider, G., Becklin, E. E., Smith, B. A., et al. 2001, AJ, 121, 525
Schneider, G., Grady, C. A., Hines, D. C. 2014, AJ, 148, 59
Schneider, G., Silverstone, M., Hines, D. C. 2006, ApJ, 650, 414
Schneider, G., Smith, B. A., Becklin, E. E., et al. 1999, ApJ, 513, L127
Schneider, G., Weinberger, A., Becklin, E. 2009, AJ, 137, 53
Schneider, G., Wood, K., Silverstone, M. D. 2003, AJ, 125, 1467
Simmons, J. F. L., & Stewart, B. G. 1985, A&A, 142, 100
Smith, B. A., & Terrile, R. 1984, Science, 226, 1421
Stark, C. S., Schneider, G., Weinberger, A. J. et al. 2014, ApJ, 789, 58
Walsh, K. J., et al. 2011, Nature, 475, 206
Weinberg, J. L., and Hahn, R. C. 1980, IAUS, 90, 19
Weinberger, A. J., Becklin, E., Schneider, G., et al. 1999, ApJ, 525, L53
Wood, K, Lada, C. J., Bjorkman, J. E., et al. 2002, ApJ, 567, 1183
Wyatt, M. C., Dermott, S., Telesco, C., et al. 1999, ApJ, 527, 918